\documentclass[onecolumn,amsmath,amssymb,nofootinbib,12pt]{article}
\usepackage{jheppub}
\usepackage{ifpdf}

\usepackage[bb=boondox]{mathalfa}

\usepackage{graphicx,subcaption}
\graphicspath{ {figures/} }
\usepackage{amsfonts}
\usepackage{amsmath}
\usepackage{amssymb}
\usepackage{epsfig}
\usepackage{pdfpages}
\usepackage{bbm}
\usepackage{graphicx,epstopdf}
\usepackage[numbers]{natbib} 
\usepackage[makeroom]{cancel}
\usepackage{hyperref}
\hypersetup{pdftex,colorlinks=true,allcolors=blue}
\usepackage{array}
\usepackage[export]{adjustbox}

\usepackage[normalem]{ulem}

\newcommand{\beq}{\begin{equation}}
\newcommand{\eeq}{\end{equation}}
\newcommand{\beqa}{\begin{eqnarray}}
\newcommand{\eeqa}{\end{eqnarray}}

\newcommand{\beqs}{\begin{equation}\begin{aligned}}
\newcommand{\eeqs}{\end{aligned}\end{equation}}

\newcommand{\bvphi}{{\bar\varphi}}

\newcommand{\dd}{\text{d}}
\DeclareMathOperator{\arcsinh}{arcsinh}
\DeclareMathOperator{\arccosh}{arccosh}

\newcommand{\eg}{{\it e.g.,}\ }
\newcommand{\ie}{{\it i.e.,}\ }
\newcommand{\mt}[1]{\textrm{\tiny #1}}
\newcommand{\reef}[1]{(\ref{#1})}
\newcommand{\ssc}{\scriptscriptstyle}

\newcommand{\B}{\mathcal{B}}
\newcommand{\bR}{\mathbf{R}}
\newcommand{\veps}{\varepsilon}

\title{\Large Quantum Extremal Islands Made Easy, Part IV:\\ Massive Black Holes on the Brane}

\author[a,b,c]{Guglielmo Grimaldi}
\author[a,b,d,e]{Juan Hernandez}
\author[a]{and Robert C. Myers}

\affiliation[a]{Perimeter Institute for Theoretical Physics, Waterloo, ON N2L 2Y5, Canada}
\affiliation[b]{Dept. of Physics \& Astronomy, University of Waterloo, Waterloo, ON N2L 3G1, Canada}
\affiliation[c]{Martin Fisher School of Physics, Brandeis University, Waltham MA 02453, USA}
	\affiliation[d]{Theoretische Natuurkunde, Vrije Universiteit Brussel, Pleinlaan 2, B-1050 Brussels}
	\affiliation[e]{The International Solvay Institutes, Pleinlaan 2, B-1050 Brussels, Belgium}

\emailAdd{ggrimaldi@brandeis.edu}
\emailAdd{rmyers@perimeterinstitute.ca}
\emailAdd{juan.hernandez@vub.be}

\preprint{BRX-TH-6696}

\abstract{We study two-dimensional eternal black holes with non-zero mass, where each asymptotic boundary is in contact with a CFT on a circle, following the doubly holographic braneworld models constructed in \cite{parti,partii,partiii}. We compute the Page curve of the black hole (or the bath CFTs), which amounts to finding different geodesics in the bulk BTZ geometry with a Randall-Sundrum brane falling into the black hole. We also explore the possibility of including an intrinsic JT gravity action on the brane. As expected,  the generalized entropy rises linearly at early times. However, there is a transition to a late-time phase in which the entropy remains constant. The value of the late-time entropy depends on the size of the thermal baths. For a small size, it corresponds to the thermal entropy of the baths, while for large size, it corresponds to twice the horizon entropy of the black hole. The critical size and the Page time are proportional to ratio of the central charges of the conformal defect and the bath CFT.}

\begin{document} 
\maketitle
\flushbottom

\section{Introduction}\label{sec:1}

Reconciling Hawking's calculation of black hole evaporation \cite{Hawking:1974rv,hawkingradiation,Hawking76} with the idea that quantum gravity is unitary has been a longstanding puzzle, \eg see \cite{Polchinski:2016hrw,Harlow_2016,Mathur:2009hf}. However,  remarkable recent progress has made it possible to compute the Page curve in a controlled manner \cite{Almheiri_2019,penington}. This progress has led to an explosion of explorations of the information paradox and the Page curve \cite{Chen:2019uhq,Chen:2020jvn,Almheiri:2019psy,Marolf:2020xie,almheiri2019eternal,Chen:2019iro,Gautason:2020tmk,Anegawa:2020ezn,Balasubramanian:2020hfs,Hartman:2020swn,Hashimoto:2020cas,Hollowood:2020cou,Alishahiha:2020qza,Geng:2020qvw,Li:2020ceg,Chandrasekaran:2020qtn,Almheiri:2020cfm,Bak:2020enw,Hollowood:2020kvk,Ling:2020laa,Marolf:2020rpm,Goto:2020wnk,Basak:2020aaa,Caceres:2020jcn,Karananas:2020fwx,Balasubramanian:2020coy,Balasubramanian:2021wgd,Uhlemann:2021nhu,Neuenfeld:2021bsb,Bousso:2021sji,Verheijden:2021yrb,Geng:2021wcq,Anderson:2020vwi,Wang:2021mqq,Bhattacharya:2021jrn,Hollowood:2021nlo,Miyata:2021ncm,Ghosh:2021axl,Balasubramanian:2021xcm, Omidi:2021opl, Ahn:2021chg}.\footnote{This progress has also stimulated new insights from a variety of different perspectives \cite{Penington:2019kki,Rozali:2019day,Bousso:2019ykv,Akers:2019nfi,Sully:2020pza,Chen:2020wiq,Giddings:2020yes,Kim:2020cds,Verlinde:2020upt,Liu:2020gnp,Harlow:2020bee,Nomura:2020ska,Hsin:2020mfa,Gaddam:2020mwe,Pasterski:2020xvn,Raju:2020smc,Bousso:2020kmy,Cheng:2020vzw,Su:2021lll,Flanagan:2021ojq,Geng:2021iyq,Azarnia:2021uch, Gaddam:2021zka}.} 

The new insights are summarized by a general rule, known as the `island rule' \cite{Almheiri_2020}. This rule states that the entanglement entropy of Hawking radiation should be evaluated as
\begin{equation}\label{QEI}
    S(\mathbf{R}) = \min\left\{\underset{\text{islands}}{\text{ext}}\left(S(\mathbf{R} \cup \text{islands})+\frac{A(\partial(\text{islands))}}{4G}\right)\right\}\,,
\end{equation}
where the radiation is captured in a (non-gravitating) subsystem or subregion $\mathbf{R}$. However, this expression also allows for contributions from
(disconnected) regions in the gravitating spacetime, called \textit{islands}. The islands contribute in the first term through the entanglement of quantum excitations in these regions with the Hawking radiation, and also with  a gravitational entropy term, \ie the Bekenstein-Hawking entropy evaluated on the boundary of the islands. The idea is that at late stages of the evaporation (or evolution in the following), the quantum entanglement becomes large enough to compete with the classical geometric contribution and the extremization procedure in eq.~\reef{QEI} may produce new saddle points, with nontrivial island regions. The striking result is that an island configuration minimizes the entropy after the Page time, but the island shrinks (to zero) as the black hole continues to evaporate, thus recovering the expected Page curve \cite{Page1,Page2}
-- see the review in \cite{almheiri2020entropy}.

The island rule was originally devised by studying a doubly holographic model of two-dimensional black holes in Jackiw-Teitelboim (JT) gravity \cite{Almheiri_2020}. Similar doubly holographic models were extensively developed in
\cite{Rozali:2019day,Geng:2020qvw,Geng:2020fxl,parti, partii,Neuenfeld:2021bsb}. These models involve introducing a tensionful brane in the AdS bulk, which backreacts the spacetime geometry and produces new graviton modes localized in the vicinity of the brane \cite{Randall1,Randall2,KarchRandall1, KarchRandall2}.
The interesting feature of these models is that the physics can be described with three different points of view: (i) the boundary and (ii) bulk perspective, as expected by the usual AdS/CFT correspondence,  and also (iii) the brane perspective, with an effective gravitational theory on the brane coupled to the boundary CFT \cite{parti,partii}.
In investigating the island formula \reef{QEI}, the real advantage of the doubly holographic models is that the computations of the entanglement entropy are purely geometric using the standard RT/HRT prescription \cite{RT,Ryu:2006ef,HRT,RangamaniBook} for holographic entanglement entropy.

The present work is a natural continuation of  \cite{partii}, which studied the island rule \reef{QEI} in the context of a doubly holographic construction built on massless hyperbolic black holes in an AdS$_{d+1}$ bulk. Investigating the scenario first studied in \cite{almheiri2019eternal}, the Page curve was recovered for an eternal black hole in equilibrium with a finite-temperature bath region coupled to each asymptotic boundary. In this context, the temperature of the bath was fine-tuned to match the curvature scale of the boundary, and the horizons were infinite in extent. In the present work, we evade both of these restrictions. However, to do so, our doubly holographic model is restricted to a $d=2$ boundary theory or a three-dimensional bulk.\footnote{A somewhat related construction appeared in \cite{JTandBTZ}, where authors studied $d=2$ black holes in JT gravity by performing a partial dimensional reduction of a BTZ geometry.}  Our results can be described from three different perspectives of our doubly holographic construction:
\begin{itemize}
\item From the {\it boundary perspective}, we have two bath CFTs each coupled to a conformal defect. These two-dimensional CFTs are each placed on a cylinder with circumference $2\pi R$, and they are entangled in a thermofield double state characterized by a temperature $T$. As the system evolves in time, the  entanglement entropy between the defect and the bath grows until it saturates to either the thermal entropy of the bath or the defect entropy. The latter is determined by the relative information capacity of the two subsystems, \ie which of the two entropies is smaller. 
\item From the {\it bulk perspective}, we have a three-dimensional eternal BTZ black hole \reef{eq:BTZmetric} extended by the inclusion of a backreacting two-dimensonal brane which extends between the two asymptotic boundaries. We consider separately the cases of ordinary tensionful branes and branes supporting JT gravity. This perspective illustrates the advantage of the doubly holographic models because the entanglement entropy is evaluated geometrically using the usual RT/HRT prescription, and we must choose the minimal surface among three different classes of extremal surfaces.
\item From the {\it brane perspective}, the bath CFTs are coupled to either side of a two-dimensional eternal black hole, as illustrated in figure \ref{fig:branepenrose}. It is this perspective which poses a potential information paradox \cite{almheiri2019eternal}. While the system is prepared in a Hartle-Hawking state and the bath CFTs are in thermal equilibrium with the gravitating region on the brane, the entanglement between the two subsystems grows as they exchange quanta at the microscopic level. 
However, the entanglement entropy saturates with the formation of a quantum extremal island in the brane in the regime where the black hole (or the defect) has the smaller information capacity.  In the regime where the thermal entropy of the bath CFTs is smaller, the entanglement entropy is saturated at this level instead. 
\end{itemize}

The remainder of our paper is organized as follows: In section \ref{sec:chapter3}, we extend the construction presented in \cite{parti,partii} to formulate a doubly holographic model for an eternal black hole with an arbitrary temperature on a two-dimensional brane, using the BTZ geometry in the three-dimensional bulk. As noted above, we consider two approaches, the first with an ordinary tensionful brane and the second with a brane which supports JT gravity.  In section \ref{sec:chapter5}, we show that, as expected, the potential information paradox is evaded by the formation of quantum extremal islands for each of these models.  We conclude with a discussion of our main results and future directions in section \ref{sec:chapter6}. We leave some details in appendices. Appendix \ref{app:Gravity} reviews the effective gravitational theory on the brane, both with and without a JT gravity term. In appendix \ref{app:geod}, we provide some expressions for the length of geodesics in $\text{AdS}_3$, which are useful for evaluating the holographic entanglement entropy in section \ref{sec:chapter5}. We examine the black hole thermodynamics and the Hawking-Page transition for our doubly holographic black holes in Appendix \ref{sec:Appendix B}.


\section{Braneworld black holes}  \label{sec:chapter3}

Refs.~\cite{parti, partii} extensively studied quantum extremal islands and the Page curve in the context of doubly holographic models -- see also \cite{Chen:2019uhq,Chen:2020jvn,Almheiri:2019psy,Rozali:2019day,Chandrasekaran:2020qtn,Bak:2020enw,Ling:2020laa,Basak:2020aaa,Caceres:2020jcn,Balasubramanian:2020hfs,Uhlemann:2021nhu,Geng:2020qvw,Geng:2020fxl,Geng:2021hlu,Geng:2021wcq,Anderson:2020vwi,Neuenfeld:2021bsb,Balasubramanian:2021xcm,Bhattacharya:2021jrn,Verheijden:2021yrb}. While their analysis  considered these phenomena in any number of dimensions, it focused on a construction involving zero-energy black holes in the bulk spacetime and on the brane. That is, in all cases, the underlying geometry was simply the AdS spacetime. This amounted to fine-tuning the temperature of the bath CFT to $T = \frac{1}{2\pi R}$, where $R$ corresponds to the curvature scale in the boundary geometry. In the following, we study a similar construction where we evade this restriction. 
However, this generalization requires us to restrict our analysis to $d = 2$, \ie a two-dimensional boundary CFT.

Following \cite{parti, partii},  our bulk gravitational theory is Einstein gravity in three dimensions coupled to a two-dimensonal brane:
\begin{equation}
\label{action}
I_{\text{bulk}} + I_{\text{brane}}= \frac{1}{16\pi G_{N}}\int \dd^{3}x\sqrt{-g}\left[R + \frac{2}{L^2}\right]  - T_0 \int \dd^2 x \sqrt{-h}\,.
\end{equation}
The parameters defining this theory include: $G_{N}$, the bulk Newton's constant; $L$, the three-dimensional AdS scale; and $T_0$, the tension of the brane.  Further, $g_{ab}$ is the metric in the three-dimensional bulk while $h_{ij}$ is the induced metric on the brane. The variation of this action \reef{action} yields the usual Einstein equation away from the brane, while the effect of the brane is accounted for by the Israel junctions conditions \cite{Israel}
\begin{equation}
\Delta K_{ij} - h_{ij}\, \Delta K^{k}{}_{k} = -8\pi G_N\,T_0\, h_{ij},
\label{eq:junction}
\end{equation}
where $\Delta K_{ij}$ is the discontinuity in the extrinsic curvature across the surface occupied by the brane. 

We begin our construction by considering the bulk geometry given by the BTZ black hole \cite{BTZ, Banados:1992gq, carlip} 
\begin{equation}\label{eq:BTZmetric}
\dd{s}^2 = -\left(\frac{r^2}{L^2} - \mu^2\right)\frac{L^2}{R^2}\dd{t}^2 + \frac{\dd{r}^2}{\frac{r^2}{L^2} - \mu^2} + r^2 \dd{\phi}^2,
\end{equation}
where  the dimensionless parameter $\mu$ is related to the mass and the horizon is positioned at $r = r_h= \mu L$.  As usual, the maximally extended geometry is holographically dual to the two copies of the boundary CFT entangled  in a thermofield double state \cite{Maldacena_Eternal}. With the asymptotic limit $r\to\infty$, we identify the background metric for the CFTs
\begin{equation}
\dd{s}^2_{\ssc \text{CFT}} = -\dd{t}^2 + R^2\,\dd{\phi}^2\,.
\end{equation}
The angle $\phi$ has period\footnote{In the usual BTZ solution, this fixed period for $\phi$ extends throughout the bulk, but it will be allowed to vary in our brane construction below.} $2\pi$ and so the circumference of this cylindrical boundary geometry is $C=2\pi\,R$. Given the metric \reef{eq:BTZmetric}, it is straightforward to evaluate the temperature as 
\begin{equation}
T = \frac{\mu}{2\pi R}\,.
\label{hawk}
\end{equation}
The black hole mass and entropy are also easily found to be
\begin{equation}\label{eq:BTZentropy}
E = \frac{L\,\mu^2}{8 G_N\,R}=  \frac{\pi^2}{3}\,c\,R\,T^2
\qquad{\rm and}\qquad
S 
= \frac{L\pi \mu}{2G_N} = \frac{2\pi^2}{3}\,c\,R\,T\,.
\end{equation}
The final expressions are written in terms of the boundary parameters $R$, $T$ and the central charge $c$, given by \cite{brown} 
\begin{equation}
c = \frac{3L}{2G_N}\,.
\label{central}
\end{equation}
Of course, these formulae \reef{eq:BTZentropy} match the energy and entropy  expected for  a two-dimensional CFT in a thermal state (at temperature $T$ and on an interval of length $C=2\pi\,R$).

\subsection*{Brane profile} 
Following \cite{parti,partii}, we place a defect on each boundary at $\phi=0$ (with $\phi \in [-\pi,\pi]$).  In the bulk, this corresponds to having the  two-dimensional brane extend across the  Einstein-Rosen bridge and reach out to intersect the two asymptotic boundaries at  $\phi = 0$. As noted above, the brane profile  is determined by the Israel junction condition \eqref{eq:junction}. 

\begin{figure}
	\centering
	\includegraphics[scale = 0.38]{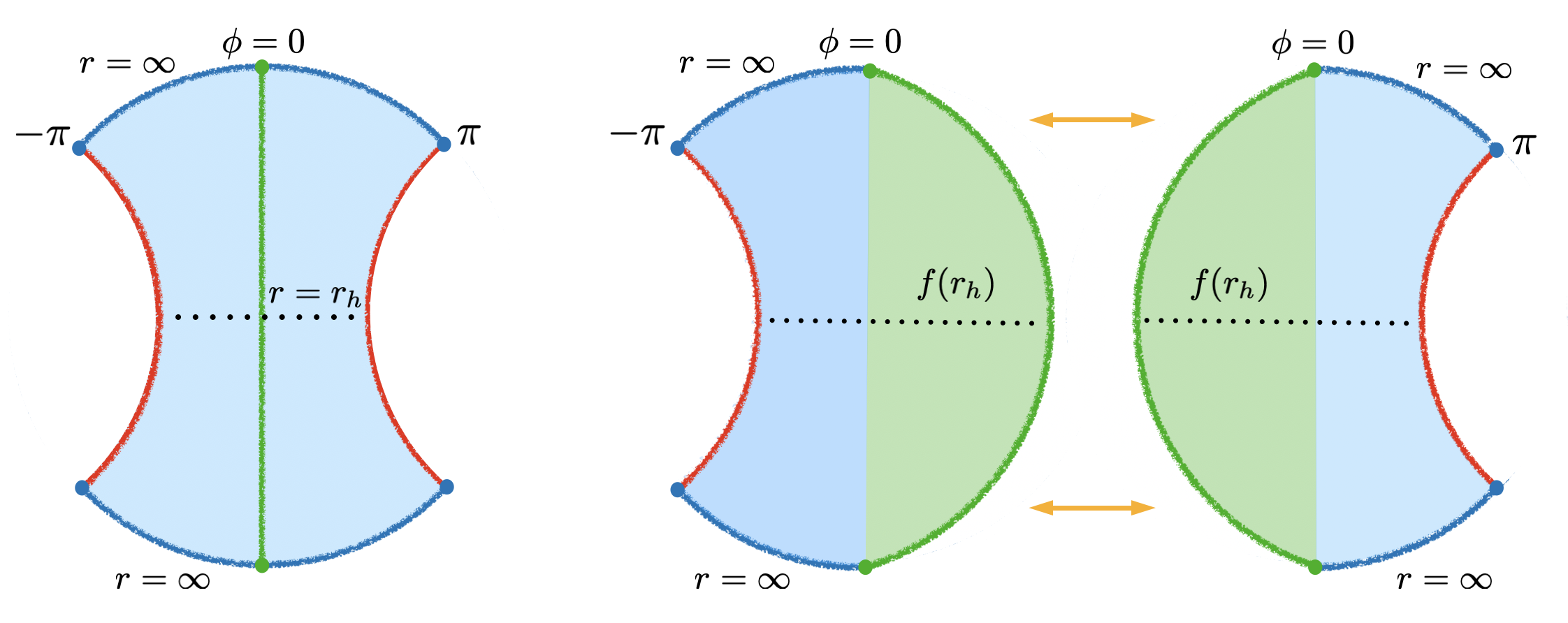}
	\caption{On the left, a $t = 0$ time slice of the BTZ spacetime with the brane shown in green, intersecting the two asymptotic boundaries at $\phi = 0$. The red lines correspond to $\phi = -\pi$ and $\phi = \pi$ and are to be identified. We show the boundaries at $r \to \infty$ in blue at the top and bottom, and the horizon $r=r_h$ with a dotted line. Recall the backreaction of the brane generates a great deal of `extra' geometry near the brane. This effect is illustrated in the sketch of the $t=0$ slice on the right, where the two regions are glued together along the green curve. The shaded green region corresponds to the extra geometry that the brane adds. }\label{fig:braneBTZ}
\end{figure}

A special feature of three dimensions is that imposing the  Einstein equations is restrictive enough to require that the bulk geometry satisfies $R_{abcd}=-1/L^2\,(g_{ac}g_{bd}-g_{ad}g_{bc})$, \ie  locally, the geometry matches that of AdS$_3$ space. As a result, inserting the two-dimensional brane does not deform the bulk geometry away from the brane surface. Further, similarly to the analysis in \cite{parti,partii}, we find that  the junction conditions \eqref{eq:junction} can be satisfied with the simple ansatz
\begin{equation}\label{eq:junction2}
K_{ij} = 4\pi G_N T_0\ h_{ij}\,.
\end{equation}
The problem then reduces to solving for an unknown static hypersurface 
\beq
\mathcal{B}\,:\quad \phi - f(r) = 0\,.
\eeq
That is, the bulk geometry on either side of the brane is described by the metric \eqref{eq:BTZmetric}, however, the angular range is now $-\pi \leq \phi \leq f(r)$ on the left side, and $-f(r)\leq \phi \leq \pi $ on the right side -- see the sketch  in figure \ref{fig:braneBTZ}.

The normal to the brane $\mathcal{B}$  is given by $n_{a} = N\,(0,-f'(r),1)$ where the normalization
$N = \big(\frac{1}{r^2} + \big(\frac{r^2}{L^2}- \mu^2\big)f'(r)^2\big)^{-1/2}$ ensures that $n_a n^a = 1$. The induced metric and extrinsic curvature for the surface can then be evaluated as
\beq
h_{ab} = g_{ab} - n_a n_b\qquad{\rm and}\qquad
K_{ab} = \frac{1}{2}\,\mathcal{L}_n g_{ab}
\eeq
where $\mathcal{L}_n$ is the Lie derivative along the normal vector. 
Then solving eq.~\reef{eq:junction2},  we find the brane profile is given by
\begin{equation}\label{eq:Brane}
\sinh{\mu\phi} = \frac{k\mu L}{r}
\end{equation}
where the constant $k$ is determined by the brane tension $T_0$
\beq \label{eq:k}
k = \frac{4\pi G_N L\, T_0}{\sqrt{1 - 16\pi^2 G_N^2L^2 \,T_0^2}}\,.
\eeq
Note that an integration constant was chosen in eq.~\reef{eq:Brane} so that,  as described above, the brane intersects the boundary at $\phi = 0$. Notice that approaching the singularity, the periodicity of $\phi$ contains a logarithmic divergence,
\beq
\Delta \phi \sim 2f(r\to0)\sim \frac{2}{\mu}\log{\frac{2k\mu L}{r}}\,.
\label{diver}
\eeq
However, the proper distance around the $\phi$ direction still contracts to zero in this limit, \ie
$\lim_{r\to0} r \Delta \phi \to 0$.
Figure \ref{fig:braneBTZ} illustrates the time slice $t=0$ for our brane construction. 

Inserting the brane profile \eqref{eq:Brane} into eq.~\eqref{eq:BTZmetric}, we find the induced metric  on the brane to be
\begin{equation}\label{eq:BH1}
\dd{s}^2_{\mathcal{B}} = - \left(\frac{r^2}{L^2}-\mu^2\right)\frac{L^2}{R^2}\dd{t}^2 + \left(\frac{1}{\frac{r^2}{L^2} - \mu^2} + \frac{k^2}{\frac{r^2}{L^2} + k^2\mu^2}\right)\dd{r}^2\,.
\end{equation}
Now with a simple coordinate transformation $r^2 = \rho^2 - k^2 \mu^2 L^2$, we obtain the following two-dimensional black hole metric on the brane
\begin{equation}\label{eq:BH2}
\dd{s}^2_{\mathcal{B}} = -\left(\frac{\rho^2}{\ell_\B^2} -\mu^2\right)\frac{\ell_\B^2}{R^2}\dd{t}^2 + \frac{\dd{\rho}^2}{\frac{\rho^2}{\ell_{\B}^2}- \mu^2}\,,
\end{equation}
where
\beq
\frac{\ell_\mathcal{B}^2}{L^2} = k^2 + 1
= \frac{1}{1 - 16\pi^2 G_N^2L^2\, T_0^2}\,.
\label{ellB}
\eeq
The above geometry \reef{eq:BH2} has a constant curvature with $\ell_\B$ being the radius of curvature, \ie the Ricci scalar is given by $\tilde{R} = - {2}/{\ell_B^2}$. Hence, the BTZ geometry in the bulk induces a locally $\text{AdS}_2$ black hole on the brane, with a horizon at $\rho = \mu\ell_B$. It is straightforward to confirm that the latter coincides with the BTZ horizon at $r=\mu L$ and that the temperature matches that found above, \ie $T=\mu/(2\pi R)$ as in eq.~\reef{hawk}. 

The energy and entropy can be evaluated  (see appendix \ref{sec:Appendix B})\footnote{Note the calculation in appendix \ref{sec:Appendix B} involves two defects on each boundary and hence the defect contribution to the entropy is twice as large there as above, \eg compare eqs.~\reef{revised} and \reef{sapp}.}
\beqa
E 
=  \frac{\pi^2}{3}\,c\,R\,T^2
\quad{\rm and}\quad
S 
&=& \frac{2\pi^2}{3}\,c\,R\,T + \frac{c}3\,\arcsinh{k}
\label{revised}\\
&=& \frac{2\pi^2}{3}c\,R\,T + \frac{c}3\,\log{2k}+\cdots\,.
\nonumber
\eeqa
In the second line, we have assumed the large $k$ regime and the ellipsis indicates higher order terms, beginning at order $1/(k\log k)$. Comparing these results with eq.~\reef{eq:BTZentropy}, the energy and entropy of a regular BTZ black hole, we see that while the energy is not modified in the defect theory, the defect makes a constant contribution to the entropy which is independent of the temperature. From the brane perspective, the defect entropy corresponds to the Wald entropy evaluated for the black hole on the brane using the  effective gravitational action -- see appendix \ref{app:Gravity}. From the boundary perspective, this contribution corresponds to the Affleck-Ludwig entropy \cite{Ludwig} associated with the defect, \ie $S_{\text{def}} = 2 \log{g}$.\footnote{Including a factor of 2 here is natural because we are considering a defect rather than a boundary CFT. With this normalization, our result matches the holographic calculation of $\log g$ in \cite{BCFT}. \label{footy99}} It is convenient to define this constant as the defect central charge, \ie
\beq
c_{\text{def}}=2\log g=\frac{c}3\,\arcsinh{k}\simeq \frac{c}3\,\log 2k\,, 
\label{cdef}
\eeq
as it characterizes the number of degrees of freedom associated with the defect.

To close here, let us note that generally we are interested in the situation where the tension approaches the critical limit $4\pi G_N L\,T_{\rm crit}=1$. That is, we consider
\beq
\veps\equiv1-4\pi G_N L\,T_{0}\ll1\,.
\label{veps}
\eeq
Hence with
\beq
\frac{\ell_\B^2}{L^2}=\frac1{\veps(2-\veps)}
\qquad{\rm and}\qquad
k=\frac{1-\veps}{\sqrt{\veps(2-\veps)}}\,,
\label{ramble}
\eeq
we are working in the regime where $\ell_\B\gg L$ and $k\gg1$.  From the bulk perspective, the brane moves far for the center of the AdS geometry in this limit, \eg $f(r)\gg1$. From the boundary perspective, this limit corresponds to a large number of degrees of freedom on the defect, \ie $c_{\text{def}}\gg1$. Further,  the ratio of the central charges associated with the defect and the CFT is large, \ie $c_{\text{def}}/c\sim \log 2k \gg 1$.
It was argued in \cite{Rozali:2019day} that the information leaks slowly from the defect to the bath in this regime. 

\subsection*{JT branes}
In recent studies of the Page curve \eg \cite{Almheiri_2019,Almheiri_2020,almheiri2019eternal}, there has been great interest in studying black holes in two-dimensional Jackiw-Teitelboim (JT) gravity \cite{jackiw,teitelboim}. Hence it is natural to extend our work here to consider branes which support JT gravity.
In this case, the brane action $I_\mt{brane}$ in eq.~\reef{action} is replaced with
\beq
I_{\rm JT} = \frac{1}{16\pi G_{\rm brane}} \int d^2x \sqrt{-h} \left[ \varphi_0 \tilde{R} + \varphi \left(\tilde{R} + \frac{2}{\ell_{\rm JT}^2} \right)\right]\,.\label{JTbrane20}
\eeq
The interested reader will find a more detailed discussion in appendix \ref{app:Gravity}, however, the key difference is that the brane geometry is now governed by the dilaton equation:  $\tilde{R} = -2/\ell_{\rm JT}^2$. Hence as before, the brane geometry corresponds to that of AdS$_2$ with $\ell_\B=\ell_{\rm JT}$. Therefore, the brane profile and induced metric are again described by eqs.~\reef{eq:Brane} and \reef{eq:BH2} where
\beq
\frac{\ell_\mathcal{B}^2}{L^2} = k^2 + 1
= \frac{\ell_{\rm JT}^2}{L^2} \,.
\label{ellB2}
\eeq

Of course, the dilaton plays a central role in determining the dynamics of JT gravity \cite{Maldacena:2016upp}, and in the present case, this scalar field makes an essential contribution to the horizon entropy on the brane. Hence, as well as the brane geometry, we must also consider the dilaton profile on the brane. While $\varphi_0$ is simply a (large) constant, $\varphi$ grows outside of the horizon. The total dilaton profile is then given by
\beq
\varphi_0+\varphi\simeq\bvphi_0+\bvphi=\bvphi_0+ \frac{\bvphi_r}{\ell_\B\,\mu} \,\rho\,.
\label{profileA}
\eeq
As described in appendix \ref{app:Gravity}, the bare dilaton $\varphi_0+\varphi$ differs from the shifted scalar  $\bvphi_0+\bvphi$ by a constant of order $1/k^2$.

The energy and entropy are given by (see eqs.~\reef{ener88} and \reef{entro88}\footnote{In comparing the entropy here with eq.~\reef{entro88}, remember there is a single defect in each boundary CFT here while there are two defects in each boundary for the system studied in appendix \ref{sec:Appendix B}.})
\beq
E = \frac{\pi^2}{3}\,c\,  R T^2\,, \qquad
S = \frac{2\pi^2}{3}\,c\,RT +  c'_{\rm def}\,,
\label{thermo88}
\eeq
where $c'_{\rm def}$ is the modified central charge of the defect dual to the brane supporting JT gravity,\footnote{As indicated here, the dilaton contribution also appears in the holographic calculation of the  Affleck-Ludwig entropy associated with the defect \cite{BCFT}.} \ie
\beq
c'_{\rm def}=2\log g'=\frac{c}3\,\arcsinh k + \frac{\varphi_0+\bar{\varphi}_r}{ 4G_{\rm brane}}\,.
\label{cdef2}
\eeq
The increase coming from the second term is simply contribution to the  horizon entropy coming from JT gravity on the brane.

\begin{figure}[h]
	\centering
	\includegraphics[scale = 0.75]{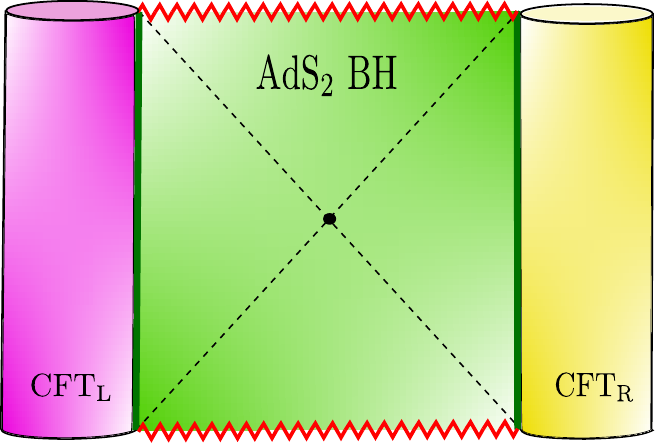}
	\caption{Penrose diagram of the system from the brane perspective.  The brane inherits an eternal black hole geometry \reef{eq:BH2}. The two bath CFTs are defined on a cylinder and are coupled to the gravity theory on  the brane through the green junctions at $\phi = 0$. }
	\label{fig:branepenrose}
\end{figure}

\vspace{1em}
To conclude here, we summarize our construction from the point of view of the three different holographic perspectives: In the bulk perspective, the system is completely geometric and is described by a (locally) BTZ geometry with an $\text{AdS}_2$ brane spanning the Einstein-Rosen bridge. The boundary perspective describes the system as two CFTs each coupled to a conformal defect and defined on a cylinder of size $2\pi R$. These two theories are then entangled in a thermofield double state with temperature $T=\mu/2\pi R$. From the {brane perspective}, we have a two-dimensional black hole on the brane that is coupled to the two boundary CFTs. The Penrose diagram for this setup in the brane perspective  is shown in figure \ref{fig:branepenrose}.


\section{Page curve}  \label{sec:chapter5}

In this section, we use our construction to examine the Page curve for an eternal black hole and two nongravitating baths, following the discussion in \cite{almheiri2019eternal}. More precisely, as noted above, the brane perspective of our system describes an eternal $\text{AdS}_2$ black hole \reef{eq:BH2} coupled (on each side) to a bath CFT on a cylinder $\mathbb{R}\times S^1$ of radius $R$ -- see figure \ref{fig:branepenrose}. The quantum fields on the $t=0$ slice are prepared in the Hartle-Hawking state with a Euclidean path integral. An information paradox then arises as follows: The black hole and the two baths continuously exchange thermal quanta, increasing the entanglement entropy between the gravitating region and the bath. In particular, quanta from the bath regions may fall into the gravitational region on the brane and disappear behind the black hole horizon. Similarly, Hawking quanta emitted by the black hole may escape the gravitating brane and be absorbed by the bath. The paradox stands in the fact that if the black hole is a quantum system with a finite number of degrees of freedom then the final entanglement entropy must be limited by this number. Of course, the excess entanglement entropy can be avoided by the formation of island in eq.~\reef{QEI} and in which case the paradox is evaded. 

This discussion was originally formulated in context of two baths of infinite size, in which case the bath had the capacity to hold an infinite amount of information. In the present case, both of the boundary baths have a finite size and so have a finite number of degrees of freedom. We will see that the finite information capacity of the bath will be the limiting factor in the growth of the entanglement entropy for small baths. In this case, no quantum extremal island forms and the details of the gravitational system on the brane are not important. Instead, the entanglement entropy is saturated at the thermal entropy of the two bath regions, as observed in the early work of \cite{Hartman_2013}. A similar saturation of entropy without islands was observed for finite intervals of semi infinite baths in~\cite{Balasubramanian:2021xcm}. As we increase the size of the bath regions, their information capacity eventually surpasses that of the black hole and  the initial growth of the entanglement is saturated by the black hole entropy, as observed in  \cite{almheiri2019eternal}.

To formulate a precise calculation, we exclude two small regions around the defects in each bath, \ie $\overline{\mathbf R}_{L,R} = \{\phi: -\phi_{\Sigma}<\phi<\phi_{\Sigma}\}$ where $\phi_{\Sigma}\ll1$. We then consider the entanglement entropy of the combination of the bath regions $\mathbf{R} = \mathbf{R}_R\cup\mathbf{R}_L$ where $\mathbf{R}_{L,R} = \{\phi: |\phi|>\phi_{\Sigma}\}$ in the two boundary CFTs. This is the region where the Hawking radiation from the gravitating brane accumulates. From the brane perspective, the Page curve saturates when a nontrivial quantum extremal surface forms on the brane to extremize the island rule \reef{QEI}. An advantage of our doubly holographic model is that the calculations of the entanglement entropy reduce to the usual RT/HRT prescription \cite{RT,Ryu:2006ef,HRT,RangamaniBook} from the bulk perspective. In three dimensions, the RT surfaces are simply geodesics and further, since the bulk geometry is locally identical to AdS$_3$, their `area' (\ie their length) is easily computed -- see appendix \ref{app:geod}. 
\begin{figure}[t]
	\centering
	\includegraphics[scale = 0.4]{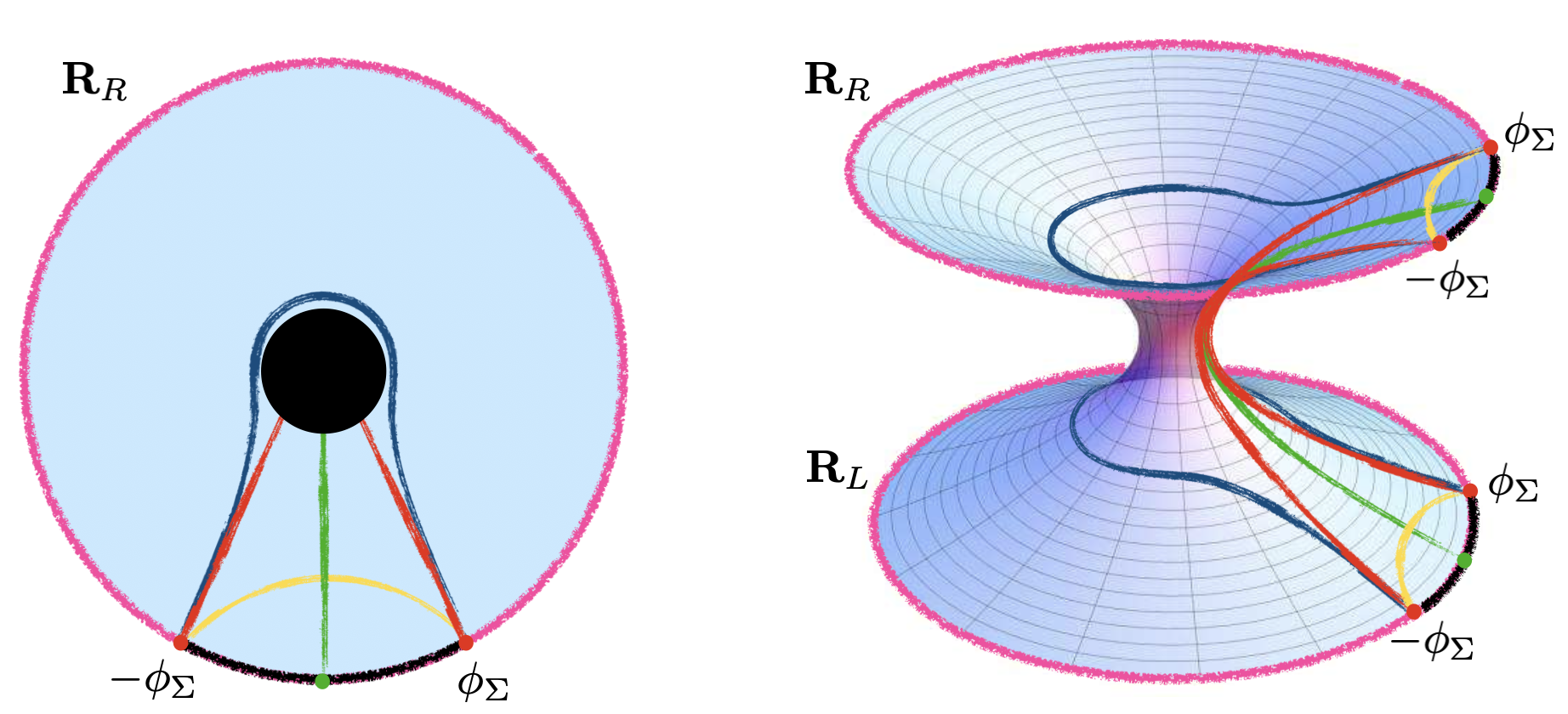}
	\caption{Sketches of the three classes of extremal surfaces appearing in calculations of the holographic entanglement entropy for the two regions $\mathbf{R} = \mathbf{R}_R\cup\mathbf{R}_L =  \{\phi: |\phi|>\phi_{\Sigma}$ in both asymptotic boundaries$\}$ (shown in magenta). On the left, a time slice of the right side of the eternal black hole. On the right, the same time slice but now the embedding diagram shows both sides of the Einstein-Rosen bridge. The brane corresponds to the green curve but recall that the backreaction of the brane generates extra geometry near the brane -- see figure \ref{fig:braneBTZ}.  In red, we have the extremal surfaces (i) contributing to the early-time phase. In blue and yellow, we have the (ii) and (iii) surfaces contributing to the thermal and island phase respectively. The latter exchange dominance depending on the size of the bath. }
	\label{fig:RTsurfaces}
\end{figure}

Of course, as is common, there may be more than one extremal surface and then one must determine which provides the minimal length and hence corresponds to the dominant saddle in the RT calculation \cite{RangamaniBook}. A valid candidate RT surface needs to be anchored to the boundaries of $\mathbf{R}=\mathbf{R}_L\cup\mathbf{R}_R$ and must be homologous to this boundary region. It is not hard to see that there are  three classes of valid candidate surfaces for the present problem. Each of these is comprised of two disconnected geodesics. As sketched in figure \ref{fig:RTsurfaces}, we have: (i) in red, two geodesics connecting points on the opposite boundaries and hence, which traverse the Einstein-Rosen bridge; (ii) in blue, two geodesics connecting points on a single boundary and which circumnavigate around the horizon, rather than crossing the brane; and (iii) in yellow, two geodesics connecting points on a single boundary and which cross the brane.

Let us make a few remarks. The surfaces (i) traverse the full geometry and will be time dependent, as will be made more clear later. This class of surface should therefore be the dominant one during the initial growth phase of the Page curve. In contrast, because surfaces (ii) and (iii) fully lie on a single side of the eternal black hole geometry, they will be geodesics lying on a time-slice of the BTZ geometry \reef{eq:BH2} and their lengths will be time-independent. Thus, we surmise that they will become dominant in the late time phase of the Page curve. Further, the surfaces (iiii) cross the brane and so they will be relevant in a regime where a quantum extremal island forms from the brane perspective. To distinguish among the three surfaces, we will denote the phases when each is dominant as (i)  the early-time or growth phase; (ii) the thermal phase; and (iii) the {island phase}. We now proceed with the computation of the entanglement entropies for the various phases and the evaluation of the Page curve.

\subsection*{Early-time phase}
We begin with the class (i) of extremal surfaces which will describe the early-time phase. Here, the RT surfaces cross the Einstein-Rosen bridge connecting the two asymptotic boundaries.  To find these geodesics, we  make use of the Kruskal extension of the BTZ spacetime, described in appendix \ref{app:geod}. In particular, a curve that crosses from boundary to boundary will necessarily have to cross the line $u = 0$ for some $v$. To find the geodesic distance, we minimize among all choices of $v$. That is, we first find the distance $d_R$ from the right boundary to $u = 0$, then the distance $d_L$ from the left boundary to $u=0$, and minimize their sum $d = d_L + d_R$ for all choices of $v$ \cite{stanford}. If the geodesics extend to some large radial cutoff $r=r_{\rm max}$, we obtain the following length 
\begin{equation}
\text{Length}(\Sigma_{\text{early}}) = 2L\log\left({\frac{2r_{\text{max}}}{\mu L}}\right) + 2L\log\left(\cosh{\frac{\mu t}{R}}\right)\,.
\end{equation}
To get the total entropy we multiply by two to account for both surfaces, hence obtaining
\begin{align}\label{eq:Searly}
S_{\text{early}}(\mathbf{R}) = \frac{\text{Length}(\Sigma_{\text{early}})}{2G_N} = \frac{2c}{3}\log\left(\frac{1}{\pi \delta T}\right) + \frac{2c}{3}\log\left(\cosh(2\pi T t)\right)\,,
\end{align}
where  the final answer is written in terms of  boundary variables: $c=3L/2G_N$, $T = \mu/2\pi R$, and the short-distance cutoff in the boundary CFT
\begin{equation}
\delta = \frac{LR}{r_{\text{max}}}\,.
\end{equation}
For times greater than the thermal scale, \ie $t\gtrsim 1/(2\pi T)$, the above result \reef{eq:Searly} reduces to
\begin{equation}
S_{\text{early}}(\mathbf{R})\simeq \frac{2c}{3}\left[\log\left(\frac{1}{2\pi \delta T}\right) +2\pi T\, t \right]\,.
\label{early0}
\end{equation}
Hence we find the linear growth expected from Hawking's calculation. This result agrees with earlier results in  the literature \cite{almheiri2019eternal, Hartman_2013, partii}. Let us note here that the rate of increase of the entanglement entropy is given by
\beq
\partial_t S_{\text{early}}(\mathbf{R})\simeq  \frac{4\pi c}{3}\,T=2\,s\,,
\label{rate}
\eeq
where $s$ corresponds to the entropy density of the thermal CFT -- compare to eq.~\reef{eq:BTZentropy}.

\subsection*{Late-time phase}
Let us now consider the late-time phase, where the entropy should remain constant. As described above, there are two classes of geodesics available to describe this phase, \ie the (ii) and (iii) surfaces. As is well known, because the BTZ geometry is static, the corresponding geodesics lie on a constant time slice bulk geometry and hence have a constant length.  Let us begin with class (ii) which circumnavigates the horizon, avoiding the brane and hence avoiding the formation of islands. The length can be evaluated using eq.~\eqref{eq:distance} taking $\Delta\phi = 2\pi - 2\phi_{\Sigma}$, $t_1 = t_2$ and $r_1=r_2=r_{\rm max}\gg L$ large $r$. The corresponding entanglement entropy is
\begin{equation}
S_{\text{thermal}}(\mathbf{R}) = \frac{\text{Length}(\Sigma_{\text{thermal}})}{2G_N} = \frac{2c}{3}\left[\log\left(\frac{1}{\pi \delta T}\right) + \log\left(\sinh(\pi T(C-\Delta \ell))\right)\right]\,,
\label{CC2}
\end{equation}
where we have introduced the circumference $C = 2\pi R$ of each bath, and the width $\Delta \ell = 2\phi_{\Sigma} R$ of the gap around the defect. As expected, the above entropy is clearly independent of time, however, the result is dependent on the size of the CFT bath. In fact, the expression in eq.~\reef{CC2} precisely matches twice the entanglement entropy of a thermal mixed state for a two-dimensional CFT at temperature $T$ and on an interval of length $C-\Delta\ell$ \cite{Calabrese:2004eu,Calabrese:2005zw}. Of course, the factor of two comes because there are two bath regions, one on each of the asymptotic boundaries.
The latter was derived for a finite interval in an infinitely long system, but it still applies for finite systems with holographic CFTs. Of course, for a large bath or high temperatures, the entropy simplifies to
\beq 
S_{\text{thermal}}(\mathbf{R}) \simeq \frac{2c}{3}\left[\log\left(\frac{1}{2\pi \delta T}\right) + \pi T(C-\Delta \ell)\right]\,.
\eeq
Here, the first term is the same UV-divergent boundary term as appears in the early-time result \reef{early0}. The second term is easily recognized as twice the thermal entropy of the CFT at temperature $T$ and on an interval of proper length $C - \Delta \ell$ -- compare to eq.~\reef{eq:BTZentropy}. Hence, the capacity of the boundary baths to store information grows linearly with their size.

We expect the above growth to stop at some large $C$ (or large $R$), when the dominant RT surface changes to another configuration. This other configuration corresponds to the class (iii) which crosses the brane, as described earlier. We determine the length of these geodesics as follows. First, consider a geodesic from the boundary point $(t_1,r_1=r_{\rm max},\phi_1= - \phi_{\Sigma})$ to a point on the brane $(t_2=t_1,r_2=r_{\text{QES}},\phi_2= \phi_{\text{QES}})$. The subscript QES signifies that the second point is a candidate for a quantum extremal surface on the brane. As this point is on the brane, the radius and angle are related by the profile \reef{eq:Brane}, \ie  $r_{\text{QES}}= k\mu L/\sinh{\mu\phi_{\text{QES}}}$. The length may again be evaluated with eq.~\eqref{eq:distance} and doubling this length, to account for the second half of the surface extending out to the asymptotic boundary on the other side of the brane, yields
\begin{equation}\label{eq:d-island}
\text{Length}(\Sigma_{\text{island}}) = 2L\log\left(\frac{2r_{\text{max}}}{L \mu}\right) + 2L\log\left(\frac{r_{\text{QES}}\cosh\mu( \phi_{\Sigma} + \phi_{\text{QES}}) - \sqrt{r_{\text{QES}}^2 - \mu^2L^2}}{\mu L}\right)\,.
\end{equation}
To find the actual RT surface, we need to minimize the above expression with respect to the position $\phi_{\text{QES}}$.\footnote{As explained in \cite{partiii}, this step corresponds to the minimization over possible islands in eq.~\reef{QEI}.} Hence, setting the derivative with respect to $\phi_{\text{QES}}$ to vanish, we find
\begin{equation}
\sinh\mu\phi_{\text{QES}} = \frac{k\sinh\mu\phi_{\Sigma}}{\sqrt{k^2 + \cosh^2\mu\phi_{\Sigma}}}\qquad{\rm and\ hence}\ \ r_{\text{QES}} =  \frac{\mu L\sqrt{k^2 + \cosh^2{(\mu\phi_{\Sigma})}}}{\sinh{\mu\phi_{\Sigma}}}\,.
\end{equation}
Note that in the large $k$ limit, this reduces to $\phi_{\text{QES}}=\phi_{\Sigma} + O(1/k^2)$ and 
$r_{\text{QES}} =  \mu L k/\sinh{\mu\phi_{\Sigma}}+O(1)$. Further, in this limit, the corresponding entropy is
\begin{align}\label{eq:S-island}
S_{\text{island}}(\mathbf{R})&= \frac{2c}{3}\left[\log\left(\frac{1}{\pi \delta T}\right) + \log{\left(2k\sinh{(\pi T \Delta\ell)}\right)}  \right]\\
&=  \frac{2c}{3}\left[\log\left(\frac{1}{\pi \delta T}\right) + \log{2k} + \log{\left(\sinh{(\pi T \Delta\ell)}\right)}  \right]\,.
\end{align}
The three contributions above correspond to the boundary term, the defect contribution, and the thermal entropy for the gap regions of width $\Delta\ell$ around the defects. 

\subsection*{Page curve}

Combining the above results, the entanglement entropy of the bath region $\mathbf{R}$ becomes
\beq
\Delta S(\mathbf{R})=  \min{(S_{\text{early}}, S_{\text{thermal}}, S_{\text{island}} )} - S_{\text{early}}(\mathbf{R};t=0)\,.
\label{totalS}
\eeq
The subtraction of $S_{\text{early}}(\mathbf{R};t=0)$ removes the UV-divergent boundary term from each of the individual entropies. As expected, we conclude that $\Delta S(\mathbf{R})$ will exhibit two phases. At early times, we have the growth phase where the class (i) surfaces dominate yielding $\Delta S(\mathbf{R})=  S_{\text{early}} - S_{\text{early}}(\mathbf{R};t=0)$. This phase, however, is capped off by a second phase where the entropy is constant. For small baths, we find this entropy to be just thermal, \ie $\Delta S(\mathbf{R})=  S_{\text{thermal}} - S_{\text{early}}(\mathbf{R};t=0)$. For large baths, the entanglement entropy is saturated by the defect entropy, \ie $\Delta S(\mathbf{R})=  S_{\text{island}} - S_{\text{early}}(\mathbf{R};t=0)$. 
\begin{figure}
	\centering
	\includegraphics[scale = 0.6]{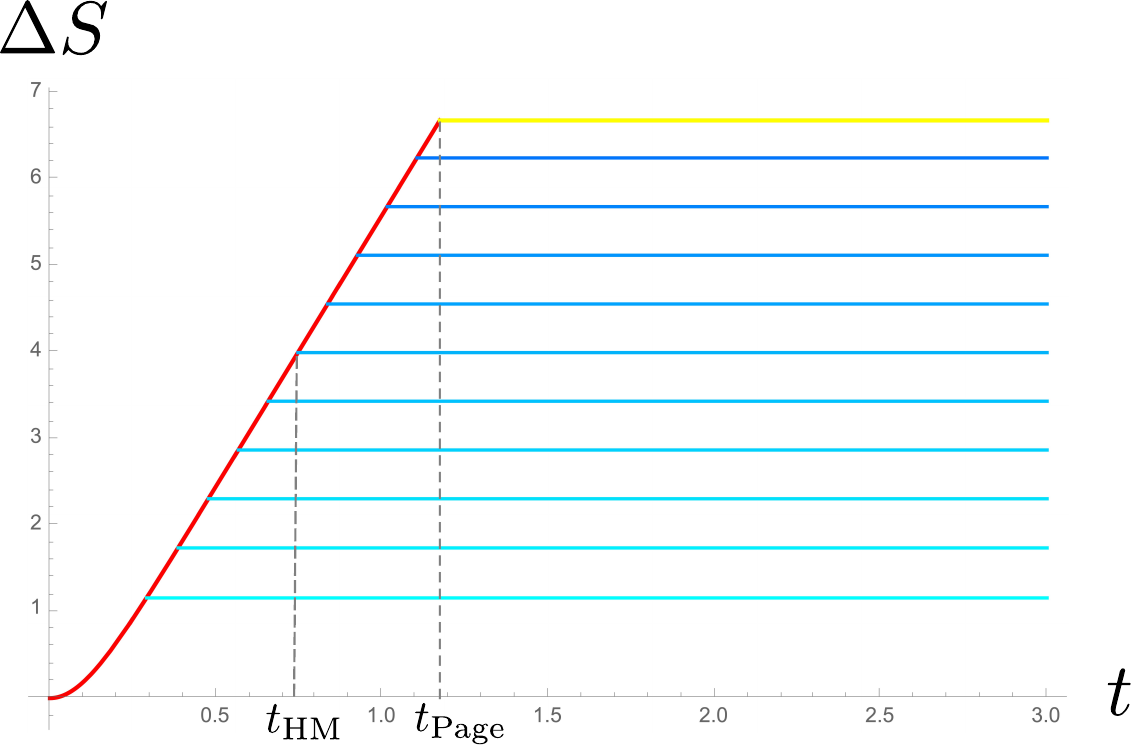}
	\caption{Plot of $\Delta S$ versus time for various bath sizes $R$, with fixed temperature $T$ (and choosing the opening angle $\Delta\phi=2\phi_\Sigma = \pi/10$ and $k=1000$). For the parameters chosen, the curves range $R = 0.1/T$ at the bottom to $R=R_{\rm crit} \approx 0.39/T$ at the top. As $R$ increases, the value at which $\Delta S$ saturates increases until it reaches the yellow curve at $R = R_{\rm crit}$. Beyond this size, the curves do not change. }
\label{needAlabel}
\end{figure}

We can find the critical size $R_{\rm crit}$ which determines the transition between the latter two possibilities by solving 
\begin{equation}
S_{\text{thermal}}(\mathbf{R}) = S_{\text{island}}(\mathbf{R})\,,
\end{equation}
which yields 
\begin{equation}
R_{\rm crit} \simeq \frac{\log 2k}{2\pi T(\pi-2\phi_\Sigma)}\,,
\label{sad22}
\end{equation}
where we have assumed $RT\phi_\Sigma\gg1$ which yields $\log\left(\sinh{(2\pi T R\phi_\Sigma)}\right)\simeq
2\pi T R\phi_\Sigma$. Of course, this assumption is consistent with the result in eq.~\reef{sad22} unless $\phi_\Sigma\sim 1/\log2k$. The behaviour of the entropy for $R<R_{\rm crit}$, \ie small bath regions, is given by
\begin{equation}
\Delta S(\mathbf{R})=  \min{(\Delta S_{\text{early}}, \Delta S_{\text{thermal}})}= 
\begin{cases}
\frac{2c}{3}\log\left(\cosh{2\pi T t}\right) & t<t_{\text{HM}}\,,\\
\frac{2c}{3}\log\left(\sinh(\pi T(C-\Delta \ell))\right) & t > t_{\text{HM}}\,.
\end{cases}
\label{options}
\end{equation}
where the transition time is given by $t_{\text{HM}} \simeq (C - \Delta\ell)/2$.\footnote{This result applies for high temperatures, \ie $T\gtrsim1/(C - \Delta\ell)$. That is, this simple expression holds when thermal wavelengths are smaller than the size of the bath regions. \label{footy88}} Here the subscript indicates  Hartman-Maldacena, since this behaviour precisely matches the results in \cite{Hartman_2013}.  We also note that the final entropy matches the thermal entropy expected for the bath region, as discussed below eq.~\reef{CC2}.

For a  large bath region with $R>R_{\rm crit}$,  the entropy follows the curve
\begin{equation}
\label{options22}
\Delta S(\mathbf{R})= \min{(\Delta S_{\text{early}}, \Delta S_{\text{island}})}  = 
\begin{cases}
\frac{2c}{3}\log\left(\cosh{2\pi T t}\right)  & t<t_{\text{Page}}\,,\\
\frac{2c}{3}\log{\left(2k\sinh{(\pi T \Delta\ell)}\right)} & t > t_{\text{Page}}\,.
\end{cases}
\end{equation}
The final entropy here matches twice the Wald entropy of the two-dimensional black hole on the brane, as shown in eq.~\reef{Wald}. We note that the above result captures contributions from an infinite series of higher curvature contributions to the gravitational action \reef{eq:induced}. The transition occurs at the Page time
\begin{equation}
t_{\text{Page}} = \frac{1}{2\pi T}\arccosh{(2k\sinh{(\pi T \Delta\ell)})} \simeq \frac{1}{2\pi T}\Big[\log 2k+\log \sinh{(\pi T \Delta\ell)})\Big]\,.
\label{options2}
\end{equation}
Of course, for large $k$, this time is dominated by the contribution proportional to $\log2k$. If we recall that $c_{\text{def}}/c\sim \log 2k$ in this regime (see the discussion after eq.~\reef{ramble}), the above result is in agreement with the discussion in \cite{Rozali:2019day} which argued that the Page time is controlled by the ratio of the central charges.


\subsection*{JT branes}

As discussed at the end of section \ref{sec:chapter3}, it is interesting to consider the case where the brane supports JT gravity -- see also appendix \ref{JTgrav}.
We consider the modifications to the Page curve for this scenario in the following.

With the addition of JT gravity \reef{JTbrane20} on the brane, the generalized entropy appearing in the island rule \reef{QEI} includes explicit gravitational entropy contributions at the edge of the island.  Hence, the position of the quantum extremal surface is found by extremizing 
\beq\label{eq:Sgen}
S_{\rm gen} = \frac{\varphi_0+\varphi}{4 G_{\rm brane}} + S_{\rm CFT}\,.
\eeq
In our doubly holographic model, the CFT contribution is given by the standard RT prescription 
\beq
S_{\rm CFT} =  \frac{\text{Length}({\Sigma_{\mathbf{R}}})}{4G_{\rm bulk}}\,,
\eeq
where $\Sigma_{\mathbf{R}}$ is the RT surface anchored at $\partial \mathbf{R}$. The gravitational contributions in eq.~\reef{eq:Sgen} appear as boundary terms in the holographic entanglement entropy because of the gravitational action on the brane \cite{parti}.
Of course, the addition of the dilaton term in eq.~\eqref{eq:Sgen} is only relevant for the class (iii) surfaces, which cross the brane. Hence JT gravity only modifies the late-time entropy for the island phase. 

\begin{figure}
	\centering
	\includegraphics[scale = 0.55]{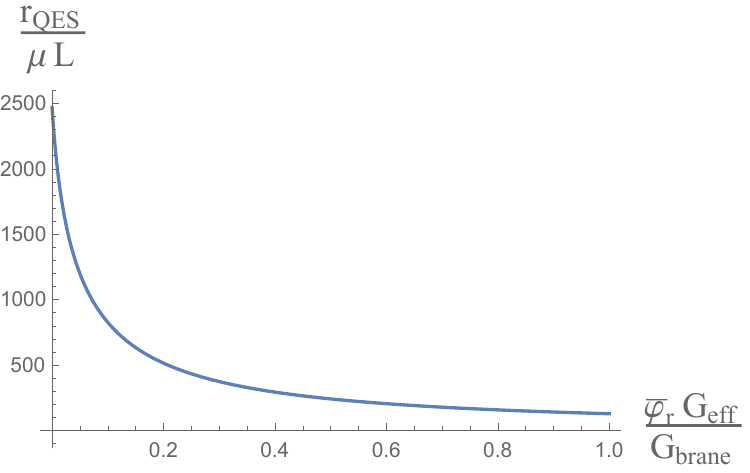}
	\includegraphics[scale = 0.55]{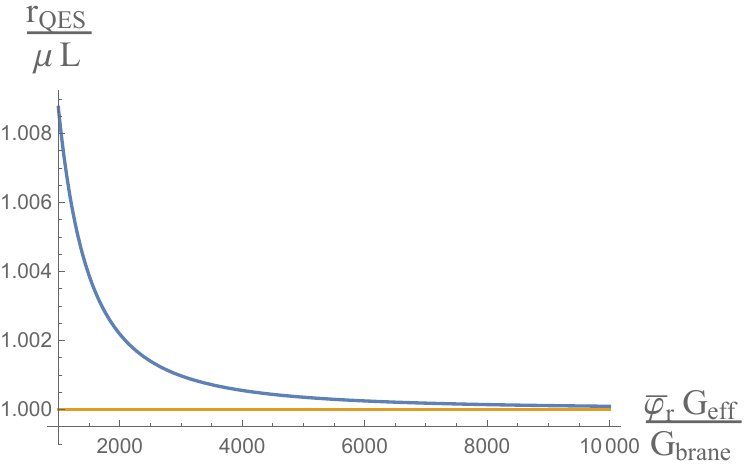}
	\caption{Change in the radial position of the quantum extremal surface for $R = 0.4$ and the same base parameters as in figure \ref{needAlabel}. 	As shown in the text, $\frac{r_{\rm QES}}{\mu L}$ is roughly inveresly proportional to $\frac{\bar{\varphi}_r G_{\rm eff}}{G_{\rm brane}}$ for $\frac{\bar{\varphi}_r G_{\rm eff}}{G_{\rm brane}} \lesssim k$ and asymptotes very close to the horizon for larger values of $\frac{\bar{\varphi}_r G_{\rm eff}}{G_{\rm brane}}$.}
	\label{fig:JTr}
\end{figure}

\begin{figure}
	\centering
	\includegraphics[scale = 0.55]{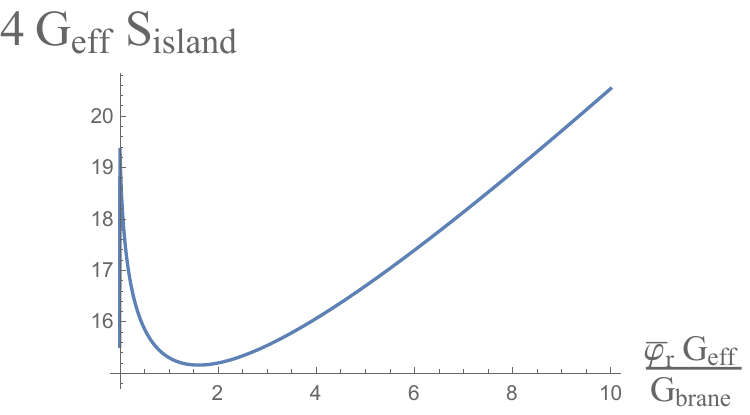}
	\includegraphics[scale = 0.55]{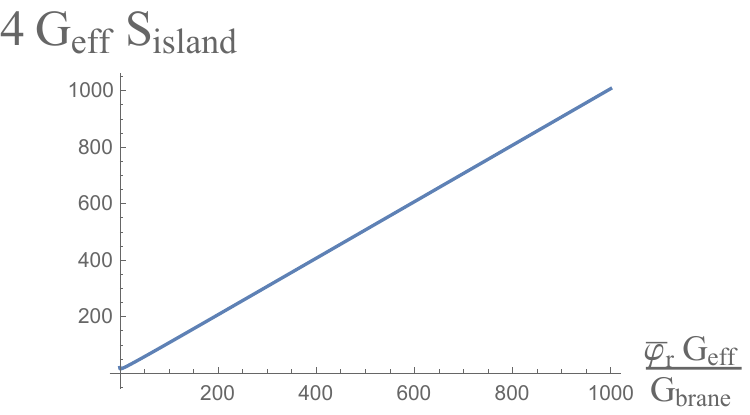}
	\caption{Change in the final entropy of the island phase for $R=0.4$ and the same base parameters as in figure ~\ref{needAlabel}. The increase quickly becomes linear.  We have also assumed $G_{\rm eff} = G_{\rm brane}$ in these plots and have ignored the $\bvphi_0$ contribution to the entropy. }
	\label{fig:JTS}
\end{figure}

The length of the candidate surfaces are given by eq.~\eqref{eq:d-island} above. Using the dilaton profile \reef{profileA}, the corresponding contribution to the entropy for these surfaces is given by
\beq
\frac{\varphi_0+\varphi}{4 G_{\rm brane}} =\frac{1}{4 G_{\rm brane}} \left(\bvphi_0+ \frac{\bvphi_r}{\ell_\B\,\mu} \,\sqrt{r^2_{\rm QES}+k^2\mu^2 L^2}\right)\,.
\eeq
As in eq.~\reef{profileA}, we have dropped a constant correction of order $1/k^2$. Combining these two expressions in the generalized entropy \reef{eq:Sgen}, we now extremize with respect to the angle $\phi_{\rm QES}$ where the surface intersects the brane
\beqs \label{eq:QES-JT0}
0 &=\frac{G_{\rm eff}}{G_{\rm brane}}\,\frac{\bar{\varphi}_r}{\sqrt{k^2+1}} \\&+ \frac{2 \sinh{(\mu\phi_{\rm QES})}\left(\cosh{(\mu \phi_{\Sigma})}\sqrt{k^2 {\rm csch}^2\left(\mu \phi_{\rm QES}\right)-1}-k\,{\rm coth}\left(\mu \phi_{\rm QES}\right)\right)}{k\,\cosh(\mu(\phi_{\rm QES}+\phi_{\Sigma}))\sqrt{k^2 {\rm csch}^2\left(\mu \phi_{\rm QES}\right)-1}+\sinh^2(\mu \phi_{\rm QES})-k^2{\rm csch}^2(\mu \phi_{\rm QES})} \,,
\eeqs
where $G_{\rm eff}=G_N/L$ is the effective Newton's constant for the gravitational theory on the brane -- see appendix \ref{JTgrav}. We note that as explained in \cite{parti}, the dilaton term produces a kink in the RT surface at the brane. That is, the RT surface is no longer smooth as it crosses the brane. The dependence of the position of the quantum extremal surface, as well as the change in the late-time entropy in the island phase, as a function of $\frac{ G_{\rm eff}}{G_{\rm brane}}\,\bar{\varphi}_r$ are shown in figures~\ref{fig:JTr} and \ref{fig:JTS}, respectively. 

From figure~\ref{fig:JTr}, we note that $r_{\rm QES}$ is a monotonically decreasing function of $\frac{ G_{\rm eff}}{G_{\rm brane}}\,\bar{\varphi}_r$. This can be made more precise as follows: For $\frac{G_{\rm eff}}{G_{\rm brane}} \bar{\varphi}_r \lesssim k$, eq.~\eqref{eq:QES-JT0} can be expanded for large $k$ keeping $\mu \phi_\Sigma$ and $\mu \phi_{\rm QES}$ finite to get

\beq
\label{eq:QES-JT}
\frac{G_{\rm eff}}{G_{\rm brane}}\,\bar{\varphi}_r + 2 \sinh{(\mu\phi_{\rm QES})}\frac{\cosh{(\mu \phi_{\Sigma})}-\cosh{(\mu \phi_{\rm QES})}}{\cosh{(\mu(\phi_{\rm QES}+\phi_{\Sigma}))}-1} + \cdots = 0\,,
\eeq
		
From eq.~\reef{eq:Brane}, decreasing $r_{\rm QES}$ corresponds to increasing $\mu\phi_{\rm QES}$, and so we examine eq.~\reef{eq:QES-JT} in the limit of large (but still smaller than $\log k$) $\mu\phi_{\rm QES}$. This yields
\beq
r_{\rm QES} \simeq  \mu L\,\frac{2k }{2\,\sinh(\mu \phi_\Sigma) +  e^{\mu \phi_\Sigma} G_{\rm eff}\,\bar{\varphi}_r/G_{\rm brane}} \,.
\label{new1}
\eeq
Therefore the radial coordinate is inversely proportional to the dimensionless parameter $\frac{G_{\rm eff}}{G_{\rm brane}} \bar{\varphi}_r$ for  $1-e^{-2\mu\phi_\Sigma} \ll \frac{G_{\rm eff}}{G_{\rm brane}} \bar{\varphi}_r \lesssim k$. However, eq.~\eqref{new1} shows that once $\frac{G_{\rm eff}}{G_{\rm brane}} \bar{\varphi}_r \sim k$, we should keep $r_{\rm QES}$ finite when taking the large $k$ limit which led to eq.~\eqref{eq:QES-JT}. Thus, taking the large $k$ limit of eq.~\eqref{eq:QES-JT0} while keeping $\mu \phi_\Sigma$ and $r_{\rm QES}$ finite, we find

\beq\label{eq:QES-JT2}
\frac{G_{\rm eff}}{G_{\rm brane}} \bar{\varphi}_r - \left(e^{\mu \phi_\Sigma} \sqrt{ \frac{r^2_{\rm QES}}{\mu^2 L^2}-1} - \frac{1}{k}\left(1+\frac{r^2_{\rm QES}}{\mu^2 L^2}\right) \right)^{-1} + \cdots = 0\,.
\eeq
The radial coordinate asymptotes to $r_{\rm QES}\to \mu L + \frac{2}{k^2} \mu L e^{-2\mu \phi_\Sigma}$ as $\frac{G_{\rm eff}}{G_{\rm brane}} \bar{\varphi}_r\to \infty$. We note that since $\mu L$ denotes the position of the horizon, the above shows that the QES approaches the horizon but remains a distance of ${\cal O} \left(\frac{\mu L}{k^2}\right)$ away. For very large $\frac{G_{\rm eff}}{G_{\rm brane}}\bar{\varphi}_r \gtrsim k$
the position of the extremal surface is
\beq
r_{\rm QES} \approx \mu L + \left(\frac{2}{k^2} + \frac{1}{2 \left(\frac{G_{\rm eff}}{G_{\rm brane}} \bvphi_r\right)^2} \right) \mu L e^{-2\mu \phi_\Sigma}\,. 
\eeq


We also note from figure \ref{fig:JTS}, the late-time entropy quickly goes over to a regime where it grows linearly with $\frac{ G_{\rm eff}}{G_{\rm brane}}\,\bar{\varphi}_r$. This behaviour arises because $r_{\rm QES}$  quickly becomes much smaller than $k\mu L$ and so by the coordinate transformation under eq.~\reef{eq:BH1}, we have $\rho= k \mu L \approx \mu \ell_B$. Hence using eqs.~\reef{ellB2} and \reef{profileA}, the dilaton at the QES saturates to the horizon value $\bvphi_0+\bvphi_r$. Hence the the final entropy simply grows linearly with $\frac{ G_{\rm eff}}{G_{\rm brane}}\,\bar{\varphi}_r$.

The (regulated) entropy in the island phase is given by
\begin{equation}
\label{eq:S-island-JT}
\Delta S_{\rm island} = 
\frac{\bvphi_0+ \bvphi_r\sqrt{1 +\alpha^2}}{2G_{\rm brane}} + \frac{2c}{3} \log{(k\,( {\rm sinh} (\pi T\Delta \ell) +\sqrt{1+\alpha^2} {\rm cosh} (\pi T\Delta \ell) -\alpha) )} \,,\end{equation}
where $\alpha = \frac{2 }{2\,\sinh(\pi T\Delta \ell) + e^{\pi T\Delta \ell}G_{\rm eff} \bvphi_r/G_{\rm brane}}$. For small values of $\frac{G_{\rm eff}}{G_{\rm brane}} \bvphi_r \ll 1- e^{-2\pi T\Delta \ell}$, we have $\alpha \approx 1/\sinh{(\pi T \Delta \ell)}$, and the entropy is given by
\beq
\Delta S_{\rm island} = \frac{\bvphi_0+ \bvphi_r \coth{(\pi T\Delta \ell)}}{2G_{\rm brane}} + \frac{2c}{3} {\rm log}\,{(2k \sinh{(\pi T\Delta \ell)})}\,, 
\eeq
which is similar to the non JT gravity case in eq.~\eqref{eq:S-island} with the additional dilaton contribution. This dilaton contribution is grater than the value of the dilaton at the horizon because $r_{\rm QES}$ is located outside of the horizon in this regime.

For large enough values of $\frac{ G_{\rm eff}}{G_{\rm brane}}\,\bar{\varphi}_r \gg e^{-\pi T \Delta \ell}$ or of $\sinh{(\pi T\Delta \ell )}\gg 1$, we have $\alpha \ll 1$ and the entropy becomes

\beq
\Delta S_{\rm island} = \frac{\bvphi_0+ \bvphi_r}{2G_{\rm brane}} + \frac{2c}{3} \left({\rm log}\,{k} + \pi T \Delta \ell \right) \,,
\eeq
and the dilaton contribution has reached its horizon value.

In this case, the transition to the island phase can be found by equating the late time limit of the early-time entropy in eq.~\eqref{eq:Searly} to the entropy in the island phase
\beq
\frac{\bvphi_0+ \bvphi_r}{2G_{\rm brane}} + \frac{1}{G_{\rm eff}} \left({\rm log}\,{k} + \pi T \Delta \ell \right) = \frac{1}{G_{\rm eff}}\left(2\pi Tt-\log{2}\right)\,,
\eeq
and so the Page time becomes
\beqs\label{eq:JT-Pagetime}
t_{\rm Page} &=  \frac{G_{\rm eff}}{G_{\rm brane}} \, \frac{\bvphi_0+\bar{\varphi}_r}{4\pi T}+ \frac{1}{2\pi T}(\log{2k} + \pi T \Delta \ell)\,.
\eeqs
Similarly, the critical size of the bath at which the island can be formed can be found by equating entropy in the island phase with the large size limit of the entropy of the thermal phase\footnote{Of course, this can also be found by matching the Page time in eq.~\eqref{eq:JT-Pagetime} and the large bath limit of the Hartman-Maldacena time
\beq
t_{\rm HM}= \frac{1}{2\pi T} {\rm arccosh}({\sinh}(2\pi R T(\pi-\phi_\Sigma))) \simeq R(\pi-\phi_\Sigma)\,.
\eeq}
\beq
\frac{\bvphi_0+\bar{\varphi}_r}{2G_{\rm brane}}  + \frac{1}{G_{\rm eff}} \left({\rm log}\,{k} + \pi T \Delta \ell \right)  = \frac{2L (\pi-\phi_\Sigma) \mu}{4G_{N}}=\frac{(\pi-\phi_\Sigma) \pi RT }{G_{\rm eff}}\,.
\eeq
The critical size at which the island becomes the final phase is therefore
\beq\label{eq:critR}
R_{\rm crit} =  \frac{1}{\pi-\phi_\Sigma} \left(  \frac{G_{\rm eff}}{G_{\rm brane}} \, \frac{\bvphi_0+\bar{\varphi}_r}{4\pi T}+ \frac{1}{2\pi T}(\log{2k} + \pi T \Delta \ell)\right)\,.
\eeq
The Page time \reef{eq:JT-Pagetime} and the critical radius \reef{eq:critR} both depend linearly on the combination $\frac{G_{\rm eff}}{2G_{\rm brane}}\,({\bvphi}_0+\bar{\varphi}_r)+ \log{2k}\sim \frac{c'_{\rm def}}{c}$ where $c'_{\rm def} = \frac{c}{3}\log{2k} + c_{\rm JT}$ is the central charge of the conformal defect, $c_{\rm JT} = \frac{1}{4} \frac{\bar{\varphi}_0+\bvphi_r}{G_{\rm brane}}$ is the central charge of the degrees of freedom dual to JT gravity, and $c$ is the central charge of the bath CFT. For a  large bath region with $R>R_{\rm crit}$,  the entropy follows the curve
\begin{equation}
\label{optionsJT}
\Delta S(\mathbf{R})= \min{(\Delta S_{\text{early}}, \Delta S_{\text{island}})}  = 
\begin{cases}
\frac{2c}{3}\log\left(\cosh{2\pi T t}\right)  & t<t_{\text{Page}}\,,\\
\Delta S_{\rm island} & t > t_{\text{Page}}\,,
\end{cases}
\end{equation}
where $S_{\rm island}$ is given in eq.~\eqref{eq:S-island-JT}.

\section{Discussion}  \label{sec:chapter6}

In our study here, we examined a doubly holographic model, which from the brane perspective, described two two-dimensional bath CFTs coupled to either side of a two-dimensional eternal black hole, as illustrated in figure \ref{fig:branepenrose}. While the bath CFTs are in thermal equilibrium with the black hole, the entanglement between the two subsystems grows as they exchange quanta at the microscopic level creating the potential for an information loss paradox \cite{almheiri2019eternal}.
Similar doubly holographic models of two-dimensional gravity were introduced in \cite{Almheiri_2020,Rozali:2019day} and extensively studied in, \eg \cite{Chen:2019uhq,Chen:2020jvn,Almheiri:2019psy,Balasubramanian:2020hfs,Geng:2020qvw,Chandrasekaran:2020qtn,Bak:2020enw,Ling:2020laa,Basak:2020aaa,Caceres:2020jcn,Uhlemann:2021nhu,Geng:2020fxl,Geng:2021hlu,Geng:2021wcq,Anderson:2020vwi,Balasubramanian:2021xcm,Bhattacharya:2021jrn,Verheijden:2021yrb,Neuenfeld:2021bsb}.  A minor difference between the present work and those earlier studies is that, for the most part, they involved $\mathbb{Z}_2$-orbifold or end-of-the-world branes analogous to those appearing in the study of holographic boundary CFTs \cite{BCFT,Fujita:2011fp,boundaryEntropy}. In contrast, following the constructions of \cite{parti,partii}, our model involved conformal defects which are dual to two-dimensional branes immersed in the three-dimensional bulk geometry.

The main distinguishing feature of our model was that the bath CFTs had a finite extent. That is, each of the two-dimensional CFTs was placed on a cylinder $\mathbb{R}\times S^1$ with circumference $2\pi R$, as illustrated in figure \ref{fig:branepenrose}. As a result, the entanglement entropy between the defect and the bath was saturated at late times, as expected for the Page curve. From the bulk perspective, the saturation is related to a transition in the RT surface giving the corresponding holographic entanglement entropy. In the early-time phase, the RT surface has two components each of which connect the two asymptotic boundaries stretching across the Einstein-Rosen bridge. In the late-time phase, the components comprising the RT surface are restricted to either of the exterior regions outside of the event horizon. However, because the baths are finite in size, there are two possibilities for the latter, as shown in figure \ref{fig:RTsurfaces}.

For large baths, as defined by eqs.~\reef{sad22} and \reef{eq:critR}, the minimal late-time RT surfaces connect the boundary points by crossing the brane. From the brane perspective, this corresponds to the formation of a quantum extremal island in the gravitating region, as described in \cite{partii}. In the high tension regime (\ie large $k$ or small $\veps$), the late-time entropy is approximately twice the horizon entropy of the two-dimensional black hole -- see eqs.~\reef{options22} and \eqref{optionsJT}. Thus this scenario evades the the information loss paradox and instead recoveres the expected Page curve.

As shown in eq.~\reef{options}, a similar result holds for small baths but in this case, the late-time RT surfaces connect the boundary points by circumnavigating the black hole (and staying away from the brane).  Hence,
from the brane perspective, no quantum extremal island appears in this case. Instead, because the RT surfaces hugs the black hole horizon,  the late-time entropy is essentially given by the thermal entropy of the two bath CFTs (plus boundary contributions), as shown in eq.~\reef{CC2}. This is precisely the behaviour found in \cite{Hartman_2013}, which examined the evolution of the entanglement entropy for a similar system in the absence of any defects (or branes).\footnote{A similar ``thermalization transition'' was found in~\cite{Balasubramanian:2021xcm} when studying the entanglement of finite intervals of semi infinite baths.} We note that, for small baths, this late-time entropy is smaller than the defect/black hole entropy found in the previous case.

In fact, the feature which distinguishes the large and small bath regimes is the relative information capacity of the two subsystems. We begin by noting that the transition from the early- to late-time behaviour is enforced by subadditivity, namely, $S(\mathbf{R}_R\cup\mathbf{R}_L)\le
S(\mathbf{R}_R)+S(\mathbf{R}_L)$. That is, the linear growth of the entanglement found in the early-time phase must stop because the entanglement entropy of the combined bath regions cannot exceed the individual entropies of the two regions, $\mathbf{R}_R$ and $\mathbf{R}_L$, on either asymptotic boundary. In fact, this bound precisely matches the final entropy in the small bath regime, and the subadditivity inequality is saturated at late times.

Of course, for the large bath regime, this bound exceeds the final state entropy. However, we note that the full system is in a pure state, \ie a thermofield double state, and hence $S({\mathbf R})=S(\overline{\mathbf R})$. That is, while we specifically evaluated the bath entropy, this is precisely the same as the entropy of the two defects (and the two small belt regions surrounding the defects). If we denote the two defect regions as $\mathbf{D}_{R,L}$, we can also consider subadditivity for the two defects to find
\beq
S({\mathbf R})=S(\overline{\mathbf{R}}) =S(\mathbf{D}_R\cup\mathbf{D}_L)\le
S(\mathbf{D}_R)+S(\mathbf{D}_L)\,.
\eeq
Of course, this bound is saturated at late times in the large bath regime, with the final entropy given by sum of the entropies of the two defects.  

Hence the late-time entropy is determined by which subsystem can hold the least information, \ie if the bath entropy $S(\mathbf{R}_R)=S(\mathbf{R}_L)$ is smaller, the baths set the bound, or if $S(\mathbf{D}_R)=S(\mathbf{D}_L)$ is smaller, the defects set the bound. Of course, if the bath is infinite in size (\eg as in \cite{almheiri2019eternal,partii}), it has infinite capacity and the final entropy is always set by the defect. The same reasoning applies in the absence of defects. That is, one could consider a thermofield double state of two CFTs on geometries of finite size. Then dividing each of the boundaries into two regions and examining the entanglement entropy of a pair of regions, one would expect to find an early-time growth phase and a late-time phase where the entropy is saturated. However, the final entropy would be given by the thermal entropies of the smaller of the two boundary pairs.\footnote{In \cite{Hartman_2013}, the boundary geometry was infinite in size and so these considerations were not needed.}

An equivalent analysis can be done in terms of the mutual information of the bath intervals and defects
\beqs
I(\mathbf{R}_R, \mathbf{R}_L) & = S(\mathbf{R}_R) + S(\mathbf{R}_L) - S(\mathbf{R}_R \cup S(\mathbf{R}_L))\,,\\
I(\mathbf{D}_R, \mathbf{D}_L) &= S(\mathbf{D}_R) + S(\mathbf{D}_L) - S(\mathbf{D}_R \cup S(\mathbf{D}_L))\,.
\eeqs
In the large bath regime, the mutual information of the two defects $I(\mathbf{D}_R, \mathbf{D}_L)$ vanishes at the Page time and the late time phase corresponds to an island configuration, while in the small bath regime, it is the mutual information of the baths $I(\mathbf{R}_R, \mathbf{R}_L)$ that vanishes at the Hartman-Maldacena time. This provides a sharp realization of the idea in~\cite{VanRaamsdonk:2010pw} that mutual information provides a measure of connectivity of the entanglement wedge in holographic models. In fact, the island is precisely what separates the entanglement wedge of $\mathbf{D}_R \cup \mathbf{D}_L$ in the brane perspective after the Page time in the large bath regime. On the other hand, the absence of an island in the small bath regime provides an indication that the entanglement wedge of $\mathbf{R}_R \cup \mathbf{R}_L$ is disconnected.

As noted above and in section \ref{sec:chapter5}, there are always three classes of extremal surfaces at any boundary time. As usual, the RT surface determining the holographic entanglement entropy is the surface with the minimal area. As discussed above, the class (ii) surfaces which hug the black hole horizon determine the late-time entropy for small baths. However, these surfaces also play a role  in the other regimes as well. In fact, the causal domains associated with the various extremal surfaces provide an example of the `Python's lunch'  \cite{Brown:2019rox,Engelhardt:2021mue,Engelhardt:2021qjs}.\footnote{This situation was also discussed in the context of quantum extremal islands and the Page curve in \cite{Neuenfeld:2021bsb} for doubly holographic models where the baths had infinite extent.} The key point is that the class (ii) surfaces are the outermost extremal surfaces, with respect to the bath regions, in any of the regimes considered here. As such, these surfaces define the `outer wedge', the domain of dependence of a partial Cauchy slice connecting the outermost extremal surface with the boundary
region \cite{Engelhardt:2017aux,Engelhardt:2018kcs}. 
It was argued in \cite{Engelhardt:2021mue,Engelhardt:2021qjs} that bulk operators with the outer wedge  admit a simple boundary reconstruction and while those outside do not, (\ie which have a complex encoding in the boundary state). In particular then, from the brane perspective, operators in the quantum extremal islands must be complex. Thus the class (ii) surfaces play an important role in characterizing how information is stored in the bath system.

For the small bath regions (and high temperatures, see footnote \ref{footy88}), we noted below eq.~\reef{options} that the entropy  saturates at $t_{\text{HM}} \simeq (C - \Delta\ell)/2$. That is, the entropy saturates at approximately one-half the transit time of the bath region. This behaviour makes no reference to the branes/defects and so precisely matches the analogous result in \cite{Hartman_2013}. One might have expected that the saturation time should correspond to the full transit time but we now argue the above result has a simple physical interpretation.\footnote{This description is motivated in part by the simple quasiparticle description of quantum quenches found in \cite{Calabrese:2006rx,Calabrese:2007rg}.}

\begin{figure}
	\centering
	\includegraphics[scale = 0.8]{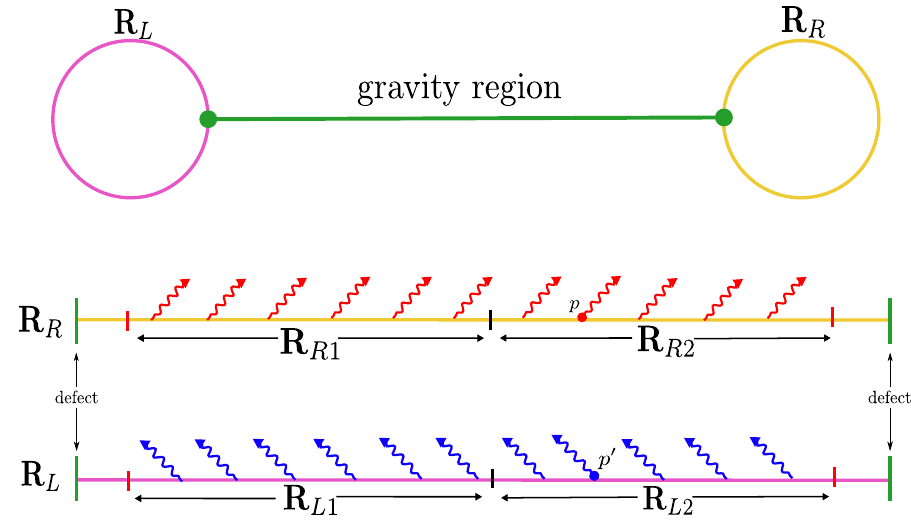}
	\caption{Picture showing the explanation given in the text. On top, we have a top view of the brane perspective, where the two boundary CFTs on the two cylinders are connected through the brane, shown in green. On the bottom, we unwrap the two CFTs keeping in mind that each respective ends are identified.}
	\label{fig:partners}
\end{figure}

Recall the description of our system from the brane perspective, as illustrated in figure \ref{fig:branepenrose}. At the top of figure \ref{fig:partners}, we show a time slice of this geometry and in the lower panel, we have unwrapped the two boundary CFTs into line segments where the endpoints connect at the defects (or junctions with the gravitating region). We also show the bath regions, $\mathbf{R}_R$ and $\mathbf{R}_L$, on these intervals in the figure. Further, we have split each of these regions into two halves (\ie, $\bR_{R1},\bR_{R2}$ and $\bR_{L1},\bR_{L2}$) as indicated.

Now the state on this $t=0$ slice corresponds to the Hartle-Hawking state \cite{almheiri2019eternal}, and so the preparation of this state involves a Euclidean path integral on a strip of width $1/2T$ connecting $\text{CFT}_R$ and $\text{CFT}_L$, as well as a hyperbolic half-disk in the gravitational region. This path integral preparation entangles the right-moving quanta in $\bR_R$ with left-moving quanta in $\bR_L$ at the same point on the corresponding cylinder,\footnote{This is a somewhat simplistic story, which should hold at high temperatures, that is, $2\pi R T \gg 1$.}  \eg right-movers in $\bR_{R1}$ are correlated with left-movers in $\bR_{L1}$. Similarly left-movers in $\bR_R$ and right-movers in $\bR_L$ are correlated, but we focus on the first case here. 

Now as we evolve forward from $t = 0$, the entanglement entropy of the bath increases because right-movers in $\bR_{R2}$ leave the bath, while the entangled left-movers in $\bR_{L2}$ remain in the bath. Similarly left-movers in $\bR_{L1}$ are leaving the bath, while their right-moving partners in $\bR_{R1}$ remain in the bath. Of course, at the same time, right-moving Hawking quanta are entering $\bR_{R1}$ and left-moving Hawking quanta are entering $\bR_{L2}$, to further increase the entropy. Further, we note that the quanta leaving the bath are quickly absorbed into the gravitational region. In particular, they do not simply traverse the junction or alternatively, the defect because of the large defect central charge.

Of course, this process continues until we reach $t=t_{\text{HM}} = (C - \Delta\ell)/2$. After this time, 
the right-movers leaving $\bR_{R2}$ and the bath are actually decreasing the entropy because their partners have already left the bath in the previous time interval. This decrease is precisely balanced by the increasing coming from the right-moving flux of Hawking quanta entering $\bR_{R1}$ at the other end. A similar story applies for the left-movers leaving $\bR_{L1}$ and the left-moving Hawking quanta entering $\bR_{L2}$. Therefore the entropy remains constant after $t=t_{\text{HM}}$. We might add that after $t=2t_{\text{HM}}$, all of the quanta that originated in the bath region at $t=0$ have left the bath. However, the entropy remains saturated because increase created by the flux of new quanta entering the bath is still balanced by the flux of quanta exiting since both sets of quanta are entangled with partners inside the black hole on the brane (or alternatively, in the defect).

The simple picture described above also explains the linear growth shown in figure \ref{needAlabel}, for which the slope was evaluated in eq.~\reef{rate} to be twice the entropy density. Both boundaries have a uniform density of (left- and right-moving) excitations, and hence there is a uniform flux of quanta leaving (and entering) either end of the bath regions. As described above, the increasing entropy arises from quanta leaving the bath regions during the growth phase, and  this flux contributes a constant factor of $s$ to the increase (since the quanta move at the speed of light). The factor of two then simply comes because there are two bath regions, one in each of the boundary CFTs. 

From the boundary perspective, a key feature characterizing the conformal defects is the defect central charge \reef{cdef}. As explained above eq.~\reef{cdef}, this central charge can be interpreted the Affleck-Ludwig entropy \cite{Ludwig} appearing in calculations for entanglement entropy in boundary CFTs. It appears as the defect entropy  
in our thermodynamic calculations in appendix \ref{sec:Appendix B}. Of course, our result is in agreement (up to a factor of two, explained in footnote \ref{footy99}) with previous results in the AdS/CFT literature for both finite and zero temperature boundary CFTs \cite{BCFT,Fujita:2011fp,boundaryEntropy}. From the brane perspective, the defect entropy becomes the coarse grained entropy of the two-dimensional black hole on the brane, which can be computed from the brane effective gravity theory as in \eqref{Wald}. The second interpretation is confirmed by noticing that the contribution $\frac{2c}{3}\log{2k}$ appearing in the island entropy matches $2S_{\text{BH}}$, \ie twice the horizon entropy of the black hole on the brane. From the brane perspective, we are naturally lead to consider the large $k$ regime so that higher curvature contributions are suppressed in the effective action on the brane, as explained in appendix \ref{app:Gravity}. This regime also corresponds to the situation where  the ratio of the central charges associated with the defect and the CFT is large, \ie $c_{\text{def}}/c\sim \log 2k \gg 1$. This in turn is the regime where information leaks slowly from the defect to the bath or from the black hole on the gravitating brane to the boundary bath \cite{Rozali:2019day}. 

We also considered branes which support JT gravity. Here, the calculations can be organized in terms of $k$  in which case the brane profile \reef{eq:Brane} remains identical to that found for the ordinary tensionful brane. Of course, the relation between $k$ and the parameters defining the brane action are very different in the two situations, \ie compare eqs.~\reef{eq:k} and \reef{label4}. It is interesting that there is a simple shift of defect central charge for the JT branes, as shown in eq.~\reef{cdef88}. That is, we found that $c'_{\rm def} - c_{\rm def} = c_{\rm JT}$ where $c'_{\rm def}$ and $c_{\rm def}$ are the central charges of the boundary defects dual to the JT and tensionful branes, respectively, with the same bulk profile. The shift $c_{\rm JT}=\frac{\varphi_0+\bvphi_r}{4G_{\rm brane}}$ is precisely the additional JT contribution to the horizon entropy of the brane black hole, as shown in eq.~\reef{Wald22}. Hence introducing  a dynamical theory of gravity directly to the brane action is increasing the number of degrees of freedom associated with the dual boundary defect.

There are many open questions for possible future investigations. A key question is to investigate the analogous physical questions in doubly holographic models for massive AdS black holes with $d > 2$. We initially looked at this possibility but realized that there were no simple solutions for the profile of a $d$-dimensional brane in the background of a massive AdS$_{d+1}$ black hole. With hindsight, we can see that this arises because the brane becomes a nontrivial source for the gravitational field in higher dimensions, and deforms the background by backreacting to the geometry in a non-trivial way. An interesting step towards such higher dimensional investigation comes with the $d=3$ construction recently examined in \cite{Emparan:2020znc,Emparan:2021hyr}.

Furthermore, as presented in Appendix \ref{sec:Appendix B}, since our background geometry is an asymptotically AdS massive black hole, there is a Hawking-Page transition at low temperatures. Actually, in Appendix  \ref{sec:Appendix B}, we compute the Hawking-Page transition for a BTZ geometry with \textit{two} branes intersecting the two boundaries at $\phi = 0$ and $\phi = \pi$. The additional second brane is required to understand the geometry for the thermal AdS phase.\footnote{It is still a mystery, and an interesting future direction, to understand the phase transition for the BTZ geometry with only one brane.} The presence of the branes in the bulk geometry drastically reduces the critical temperature in the limit where the ratio of the defect and boundary central charges is large -- see eqs.~\reef{gamble2} and \reef{gamble88}.  There are several directions which remain to be explored here. For example, examining how the phase transition is modified in the case of higher dimensions. Again, the recent $d=3$ construction \cite{Emparan:2020znc,Emparan:2021hyr} is a natural framework to investigate this question. Another question would be to investigate the effect moving the position of the second defect away from $\phi=\pi$.

Other interesting directions would be to study holographic complexity in our framework, following the approach discussed in \cite{partiii} -- see also \cite{Emparan:2021hyr,Bhattacharya:2021jrn,Bhattacharya:2021nqj, prepare:2022}. Finally, let us comment that since the horizon has a finite size in our construction, it is amenable to a fuzzball description, \eg \cite{Mathur:2005zp,Guo:2021blh}. It would be interesting to understand the appearance of quantum extremal islands on the brane from this perspective.

\begin{acknowledgments}
We are happy to thank Galit Anikeeva, Ben Craps, Matthew Headrick, Mikhail Khramtsov, Dominik Neuenfeld and Nicol\'as Vald\'es-Meller for fruitful discussions and useful comments.  Research at Perimeter Institute is supported in part by the Government of Canada through the Department of Innovation, Science and Economic Development Canada and by the Province of Ontario through the Ministry of Colleges and Universities. RCM is supported in part by a Discovery Grant from the Natural Sciences and Engineering Research Council of Canada, and by funding from the BMO Financial Group.  RCM also received funding from the Simons Foundation through the ``It from Qubit'' collaboration. Research by JH is also supported by FWO-Vlaanderen projects G006918N and G012222N, and by Vrije Universiteit Brussel through the Strategic Research Program High-Energy Physics. 
\end{acknowledgments}

\appendix


\section{Gravity on the brane}\label{app:Gravity}

As discussed in the main text, the brane enlarges the geometry and as a result, new graviton modes appear localized near the brane. This allows for a description of the brane as an effective theory of two-dimensional gravity coupled to two copies of the bath CFTs -- albeit with a cutoff. The technical details on the construction of the effective theory can be found in \cite{parti, partii}. We only report the salient points for brevity. By examining the divergences the idea is to consider the bulk gravitational theory, with appropriate Gibbons-Hawking-York boundary terms, and integrate over the radial direction the on-shell action up to the brane. Schematically, the induced action on the brane takes the form
\begin{equation}
I_{\text{induced}} = 2I_{\text{diver}} + I_{\text{brane}}
\end{equation}
where $I_{\text{brane}}$ is the usual tension term for the brane introduced in eq.~\eqref{action}, and $I_{\text{diver}}$ is the divergent boundary integral obtained from the above integration procedure. After a careful analysis, one arrives at the following action for the  gravitational theory on the brane \cite{parti}\footnote{Note that we have changed the normalization for the argument of log term in eq.~\reef{eq:induced} to match the black hole entropy on the brane with the defect entropy in eq.~\reef{revised}. The freedom to make this change arises because the ambiguity in choosing the infrared scale in the log term in $I_{\text{diver}}$, \ie see eq.~(2.23) in \cite{parti}where $L$ was chosen as a convenient scale to make the argument of the log dimensionless. Alternatively, the issue is that the coefficient of the topological $\tilde R$ term in the action \reef{eq:induced} is ambiguous because for $d=2$, infrared quantum contributions already compete with the subleading terms in the FG expansion -- see the discussion in Appendix A.2 of~\cite{partiii}.} 
\begin{equation}\label{eq:induced}
I_{\text{induced}} = \frac{1}{16\pi G_{\text{eff}}} \int \dd^2 x \sqrt{-h} \left(\frac{2}{\ell_{\text{eff}} ^2} - \tilde{R}\,\log{\left(-\frac{L^2}{8}\tilde{R}\right)}+ \tilde{R} + \frac{L^2}{8}\tilde{R}^2 + \cdots \right)
\end{equation}
where $G_{\text{eff}} = G_N/L$ and 
\beq
\frac{L^2}{\ell_{\text{eff}}^2}=2\left(1- 4\pi G_N L\, T_0\right)=2\veps\,.
\label{leff}
\eeq
Further, $\tilde{R}$ is the Ricci scalar for the induced metric $h_{ij}$ on the brane. 
Recall that $\veps$ was defined in eq.~\reef{veps}, and we focus on the regime $\veps\ll1$. Therefore  $\ell_{\text{eff}} \gg L$ and we note that this limit ensures that contributions coming from the higher curvature terms in the action \reef{eq:induced} are suppressed.

Although this action with the logarithmic term is unconventional, one can use it to derive the following metric equation of motion
\beq
0=\frac{2}{\ell_{\text{eff}} ^2}  + \tilde{R} - \frac{L^2}{8}\tilde{R}^2+\cdots\,,
\eeq
where we have reduced this to a scalar equation using the geometric identity for two dimensions: $2\,\tilde{R}_{ij}=\tilde{R}\,h_{ij}$. Comparing eqs.~\reef{ellB} and \reef{leff} for the curvature scales on the brane and in the action \reef{eq:induced}, one can easily show that this equation of motion correctly produces the leading terms for $\ell_{\text{eff}}$ in an expansion in terms of $L^2/\ell_{\B}^2$.\footnote{That is, one finds 
$\frac{L^2}{\ell_{\text{eff}} ^2}=2\left(1-\sqrt{1-\frac{L^2}{\ell_{\B} ^2}}\right)=\frac{L^2}{\ell_{\B} ^2}+\frac{1}4\,\big(\frac{L^2}{\ell_{\B} ^2}\big)^2+\cdots$ for $L^2/\ell_{\B}^2\ll1$.} 

The horizon entropy can be evaluated with the Wald formula \cite{Wald_1993}, which yields
\beqa
S_{\text{Wald}}&=&\frac{1}{4G_{\text{eff}}}\left(- \log{\left(-\frac{L^2}{8}\tilde{R}\right)}  + \frac{L^2}{4}\tilde{R}+ \cdots \right)
\label{Wald}\\
&=&\frac{1}{4G_{\text{eff}}}\left(- \log{\left(\frac{L^2}{4\ell_\B^2}\right)}  -\frac12\, \frac{L^2}{\ell_\B}+ \cdots \right)=\frac{L}{2G_N}\left( \log{2k}  +\frac14\, \frac{1}{k^2}+ \cdots \right)\,.
\nonumber
\eeqa
Using $c=3L/2G_N$ as given in eq.~\reef{central}, we see the leading term above matches the leading defect entropy in eq.~\reef{revised}. However, let us add that the second term above also matches the first subleading contribution in the expansion of the full defect entropy. 

As noted above, the logarithmic term is somewhat unusual, arising in two dimensions because of the nonlocal nature of the underlying Polyakov action \cite{Polyakov:1981rd,Skenderis:1999nb}.  This contribution to the action can also be recast in a local form, as a Liouville theory. For the latter, we introduce an  auxiliary field $\Phi$ with action
\begin{equation}
I_{\text{L}} = \frac{1}{16\pi G_{\text{eff}}}\int \dd^2{x}\sqrt{-h}\left(-\frac{1}{2}(\nabla \Phi)^2 + \Phi \tilde{R} -\frac{8}{L^2}\, e^{-\Phi}\right)\,.
\label{Louis}
\end{equation}
A simple solution for the resulting equations of motion is $\Phi = \Phi_0= -\log{(-L^2\tilde{R}/8)}$. Then substituting this solution into the action \reef{Louis}, one obtains
\begin{equation}
I_{\text{L}} = \frac{1}{16\pi G_{\text{eff}}}\int \dd^2{x}\sqrt{-h}\left(-  \tilde{R}\log{\left(-\frac{L^2}{8}\tilde{R}\right)+\tilde{R} }\right)\,,
\end{equation}
which matches the leading curvature terms in eq.~\eqref{eq:induced}. 

We can also compute the entropy of the black hole on the brane using this Liouville action \reef{Louis}. Following \cite{Myers_1994}, we again use the Wald entropy formula \cite{Wald_1993} to see that the horizon entropy is given by a logarithmic contribution
\begin{equation}\label{eq:bhentropy}
S_{\text{BH}} = \frac{\Phi}{4G_{\text{eff}}} = -\frac{1}{4G_{\text{eff}}} \log{\left(-\frac{L^2\tilde{R}}{8}\right)} 
\simeq \frac{L}{2G_N}\log{2k}\,,
\end{equation}
where we have given the final result in the  limit of large $k$. Of course, this reproduces the leading term in Wald entropy evaluated in eq.~\reef{Wald}.

\subsection{JT gravity on the brane} \label{JTgrav}

It is interesting to consider branes supporting JT gravity \cite{jackiw,teitelboim,Almheiri:2014cka,Maldacena:2016upp}.  In this case, the standard brane action in eq.~\reef{action} is replaced with 
\beq
I'_{\rm brane} = I_{\rm JT} + I'_{\rm ct}\,,
\label{JTbrane1}
\eeq
where the JT gravity action is given by (as in eq.~\reef{JTbrane20})
\beq
I_{\rm JT} = \frac{1}{16\pi G_{\rm brane}} \int d^2x \sqrt{-h} \left[ \varphi_0 \tilde{R} + \varphi \left(\tilde{R} + \frac{2}{\ell_{\rm JT}^2} \right)\right]\,,
\label{JTbrane2}
\eeq
and the counterterm
\beq
I'_{\rm ct} = - \frac{1}{4\pi G_{N}L}\int \dd^2 x \sqrt{-h}\,.
\label{JTbrane3}
\eeq
The latter cancels the effective cosmological constant from the induced gravitational action -- see above, as well as refs.~\cite{parti,partii}.\footnote{Without this counter term, a cosmological constant term with be induced with ${L^2}/{\ell_{\text{eff}}^2}=2$. The only effect of this term would be the addition of an additional source term proportional to $h_{ij}$ on the right hand side of eq.~\reef{eq:dilaton}. Of course, this new term can be eliminated by a shift of the dilaton similar to that in eq.~\reef{shift}.} With the addition of the JT gravity term, the induced action~\eqref{eq:induced} becomes
\beqs
I_{\text{induced}} =&\ \frac{1}{16\pi G_{\text{eff}}} \int \dd^2 x \sqrt{-h} \left[- \tilde{R}\log{\left(-\frac{L^2}{8}\tilde{R}\right)} + \frac{L^2}{8}\tilde{R}^2 + \cdots \right]\\
& \ \ +\frac{1}{16\pi G_{\rm brane}} \int d^2x \sqrt{-h} \left[ \bar{\varphi}_0 \tilde{R} + \varphi \left(\tilde{R} + \frac{2}{\ell_{\rm JT}^2} \right)\right]\,,
\label{label5}
\eeqs
where we have grouped the topological Einstein-Hilbert terms by shifting 
\beq
\bar{\varphi}_0 = \varphi_0 + G_{\rm brane}/G_{\rm eff}\,. \label{shift0}
\eeq
Recall from above that  $G_{\rm eff}=G_N/L$.

The dilaton equation of motion imposes $\tilde{R}=-2/\ell^2_{\rm JT}$. Hence the brane geometry is locally equivalent to $AdS_2$ with the curvature scale $\ell_\B = \ell_{\rm JT}$. Thus the brane profile \reef{eq:Brane} and the induced metric \reef{eq:BH2} are unchanged but $k$ is simply given by 
\beq
k = \sqrt{\frac{\ell_{\rm JT}^2}{L^2}-1}\,.
\label{label4}
\eeq
We will  again focus on the regime where $\ell_\B=\ell_{\rm JT}\gg L$ or $k\gg1$.

The metric equation of motion on the brane becomes
\beq\label{eq:dilaton}
-\nabla_i \nabla_j \varphi + h_{ij} \left(\nabla^2\varphi - \frac{\varphi}{\ell_{\rm JT}^2} \right) = 8\pi G_{\rm brane} \,T_{ij}^{\rm CFT} = -\frac{G_{\rm brane}}{ G_{\rm eff}} \,\frac{1}{\hat\ell_{\rm eff}^2}\,h_{ij}\,,
\eeq
where
\beq
\frac{L^2}{\hat\ell_{\rm eff}^2} = 2 \left( 1-\sqrt{1-\frac{L^2}{\ell_{\rm JT}^2}}\right)\,.  
\eeq
Note that to leading order in the regime where $\ell_{\rm JT}\gg L$, we have  $\ell_{\rm eff}\simeq\hat\ell_{\rm JT}$.
The above equation \reef{eq:dilaton} can be related to the source-free dilaton equation (which usually appears in the literature, \eg \cite{Almheiri:2014cka,Maldacena:2016upp,Engelsoy:2016xyb,Almheiri:2016fws}) using the shift
\beq
\bar{\varphi} = \varphi - \frac{G_{\rm brane}}{G_{\rm eff}} \frac{\ell_{\rm JT}^2}{\hat\ell_{\rm eff}^2}\,.
\label{shift}
\eeq
Combining eqs.~\reef{shift0} and \reef{shift}, we note that to leading order 
\beq
\varphi_0+\varphi=\bar{\varphi}_0+\bar{\varphi}
-\frac{G_{\rm brane}}{4G_{\rm eff}}\,\frac{L^2}{\ell_{\rm JT}^2}+\cdots
\,.
\label{interest}
\eeq
Hence the difference between the bare dilaton $\varphi_0+\varphi$ and the shifted dilaton $\bar{\varphi}_0+\bar{\varphi}$ is simply a constant of order $1/k^2$.

Now a standard metric on AdS$_2$ is given by
\beq
ds^2=\ell_\B^2\left[-(\tilde\rho^2-1) d\tilde t^2+\frac{d\tilde\rho^2}{\tilde\rho^2-1}\right]\,,
\label{stan2}
\eeq
where a coordinate horizon appears at $\tilde\rho=1$. These coordinates are adapted to the dilaton profile given by
\beq
\bar\varphi=\bvphi_r\,\tilde\rho\,.
\label{stan3}
\eeq
That is, surfaces of constant dilaton correspond to surfaces of constant $\tilde\rho$.\footnote{Of course, there are families of dilaton profiles related by special conformal transformations \cite{Maldacena:2016upp} but they will all take the form \reef{stan3} with the appropriate choice of coordinates.} We note that the value of the dilaton on the horizon is given by the integration constant $\varphi_r$, which will play a role in evaluating the horizon entropy -- see below. Typically, the discussion of JT gravity focuses on the boundary value of the dilaton $\varphi_b=\varphi_r/\epsilon$ at the asymptotic boundary $\tilde\rho=1/\epsilon$. Of course, the latter plays an essential role in understanding the dynamics of JT gravity \cite{Maldacena:2016upp}.

The induced metric \reef{eq:BH2} on the brane is related to the above metric \reef{stan2} by the simple rescaling of the coordinates:
\beq
\tilde\rho =\frac{\rho}{\ell_\B\,\mu} \quad{\rm and}
\quad \tilde t=\frac{\mu\,t}{R}\,.
\label{rescale}
\eeq
Hence when working with the metric \reef{eq:BH2} in the main text, the dilaton profile \reef{stan3} becomes
\beq
\bvphi= \frac{\bvphi_r}{\ell_\B\,\mu} \,\rho\,,
\label{profile}
\eeq
as appears in eq.~\reef{profileA}.

We are now prepared to evaluate the horizon entropy of the black hole on the JT brane with the usual Wald formula \cite{Wald_1993}. Focusing on the leading terms in the effective action \reef{label5} yields
\beqa
S_{\text{Wald}}&=&-\frac{1}{4G_{\text{eff}}} \log{\left(-\frac{L^2}{8}\tilde{R}\right)}  + 
\frac{\bvphi_0+\varphi}{4\,G_{\rm brane}}\Big|_{\rho=\mu\ell_{\rm JT}}+ \cdots 
\nonumber\\
&=&-\frac{1}{4G_{\text{eff}}} \log{\left(\frac{L^2}{4\ell_{\rm JT}^2}\right)}  + 
\frac{1}{4\,G_{\rm brane}}\left(\bvphi_0+\frac{G_{\rm brane}}{4G_{\rm eff}}\,\frac{L^2}{\ell_{\rm JT}^2}+\bvphi_r\right)+ \cdots 
\nonumber\\
&=&\frac{L}{2G_N}\log{2k}  +\frac{\varphi_0+\bvphi_r}{4\,G_{\rm brane}}+ \cdots \,,
\label{Wald22}
\eeqa
where we have used eqs.~\reef{shift0} and \reef{shift} to produce this final result. Comparing to eq.~\reef{Wald}, we see that the black hole entropy is increased over that of the tensionful brane by a simple shift, which can be seen as the gravitational contribution coming from the (bare) JT action \reef{JTbrane2}.


\section{Geodesics in $\text{AdS}_3$ } \label{app:geod}

In section \ref{sec:chapter5}, we are interested in computing the entanglement entropy of the combination of regions $\mathbf{R} = \{\phi: |\phi|>\phi_{\Sigma}\}$ in the two boundary CFTs. These are the bath regions where the Hawking radiation accumulates and to evaluate the corresponding Page curve, we must determine the quantum extremal surfaces (QESs) which extremize the island rule \reef{QEI}. Of course, one of the advantages of the doubly holographic model which we are studying is that these calculations are purely geometric \cite{Almheiri_2020, almheiri2019eternal, parti, partii}. That is, determining the QESs in the brane perspective reduces to the usual RT prescription \cite{RT,Ryu:2006ef} for evaluating holographic entanglement entropy (with the addition of boundary terms for a gravitating brane \cite{parti}) in the bulk perspective. In the three-dimensional bulk geometry, these RT surfaces are simply geodesics. To compute their area (\ie length), we can use the formula for the length of geodesics in pure $\text{AdS}_3$, since the BTZ geometry \reef{eq:BTZmetric} is equivalent to that of $\text{AdS}_3$, up to global identifications.

In the BTZ metric \eqref{eq:BTZmetric}, the coordinates $(t,r,\phi)$ only cover one of the two asymptotic regions on either side of the eternal black hole. For the geodesic computations, it will be more convenient to use the Kruskal coordinates for the BTZ metric which smoothly cover the full two-sided geometry, \eg see \cite{carlip, stanford}. As usual, one defines the new coordinates $(u,v,\phi)$  with $u = T-X$ and $v = T+X$ where
\beq
T(r,t) = \left(\frac{r - \mu L}{r + \mu L}\right)^{1/2} \sinh\left(\frac{\mu}{R}t\right)\quad{\rm and}\quad X(r,t) = \left(\frac{r - \mu L}{r + \mu L}\right)^{1/2} \cosh\left(\frac{\mu}{R}t\right)\,.
\label{newcoord}
\eeq
The prescription above is valid for the exterior regions $r>\mu L$. For the interior regions $r< \mu L$, one chooses
\beq
T(r,t) =  \left(\frac{\mu L-r}{\mu L+r}\right)^{1/2} \cosh\left(\frac{\mu}{R}t\right)\quad{\rm and}\quad X(r,t) =  \left(\frac{ \mu L-r}{ \mu L+r}\right)^{1/2} \sinh\left(\frac{\mu}{R}t\right)\,.
\label{eq:interior}
\eeq
With these coordinates, the metric \reef{eq:BTZmetric} becomes
\begin{equation}
\dd{s}^2 = \frac{- 4L^2}{(1 + uv)^2} \dd{u}\dd{v} + \mu^2 L^2 \frac{(1-uv)^2}{(1+uv)^2} \dd{\phi}^2\,.
\end{equation}
The singularity $(r = 0)$ now appears at $uv \to 1$ and the two asymptotic boundaries ($r \to \infty$), at $uv \to -1$. Of course, the horizons are given by $uv=0$. For the maximally extended spacetime the range of $u$ and $v$ is $-\infty < u < \infty$ and $-\infty < v < \infty$. The spacetime is thus split into four quadrants depending on the signs of $u$ and $v$. The choices for $T$ and $X$ given in eqs.~\reef{newcoord} and \reef{eq:interior} correspond to $u<0,v>0$ and $u,v>0$, respectively, which coorespond to the left exterior and the future interior regions of the BTZ black hole. One can easily take care of the remaining quadrants of the eternal black hole (\ie $u>0,v<0$ and $u,v<0$) by sending $T \to -T$ and $X \to -X$.  

Following \cite{stanford} to evaluate the lengths of geodesics in this geometry, we recall that $\text{AdS}_3$ can be defined as the hypersurface in $\mathbb{R}^{2,2}$ satisfying 
\beq
-T_1^2 -T_2^2 + X_1^2 + X_2^2 = -L^2\,.
\label{embed0}
\eeq 
Now conveniently, geodesics in AdS$_3$ are straight lines if viewed using the embedding coordinates $(T_1,T_2,X_1,X_2)$, with the proper distance $d$ between two points  (primed and unprimed) given by
\begin{equation}
\cosh{\frac{d}{L}} = T_1T'_1 + T_2T_2' - X_1X'_1 - X_2X_2' \,.
\end{equation}
Now the embedding coordinates relate to the above Kruskal coordinates and the original BTZ coordinates as follows 
\begin{align}
\begin{split}
T_1 &= \frac{u + v}{1 + uv} = \frac{(r^2-\mu^2 L^2)^{1/2}}{\mu L}  \sinh{\frac{\mu t}{R}},\;\;
T_2 = \frac{1 - uv}{1 + uv}\cosh{\mu\phi} =\frac{r}{\mu L}\cosh{\mu\phi}\,,\\
X_1 &= \frac{v-u}{1 + uv} = \frac{(r^2-\mu^2 L^2)^{1/2}}{\mu L} \cosh{\frac{\mu t}{R}},\;\;
X_2 = \frac{1 - uv}{1 + uv}\sinh{\mu\phi} =\frac{r}{\mu L}\sinh{\mu\phi}\,.\\
\end{split}
\end{align}
These expressions are valid for the exterior region on the right side. To move to the left exterior region, one simply replaces $t\to t + \frac{i}{2T}$ where $T = \frac\mu{2\pi R}$ \cite{stanford}. A general formula for the geodesic length between a pair points in the BTZ coordinates, say $(t_1,r_1,\phi_1)$ to $(t_2,r_2,\phi_2)$, reads 
\begin{equation}\label{eq:distance}
\cosh{\frac{d}{L}} =\frac{r_1 r_2}{\mu^2 L^2}\,\cosh{\mu(\phi_1 - \phi_2)} - \frac{\sqrt{(r^2_1-\mu^2L^2)(r^2_2-\mu^2L^2)}}{\mu^2L^2} \,\cosh{\frac{\mu(t_1 - t_2)}{R}} \,.
\end{equation}


\section{BH thermodynamics and HP transition}\label{sec:Appendix B}

Here we examine our construction from the perspective of black hole thermodynamics. That is, we consider the solutions of the Euclidean Einstein equations as saddle points of the quantum gravity path integral. Further, the Euclidean time coordinate must be periodic for the corresponding black hole geometry to be smooth and following the standard QFT approach, the Euclidean path integral then yields the thermal partition function of the system, \eg see \cite{Witten:1998zw,Harlow_2016,Gibbons:1976ue,SWH1979}.  Of course, an interesting phenomenon in the context of holography is the phase transition discovered by Hawking and Page \cite{HawkingPage} for  asymptotically AdS black holes. That is, the dominant saddle point below a certain critical temperature  becomes thermal AdS, \ie Euclidean AdS space with a periodic time direction. According to the holographic dictionary, the AdS black hole describes a deconfined phase of the dual boundary theory (with entropy proportional to the central charge) while thermal AdS describes a confined phase (with order-one entropy). Hence the HP transition describes a deconfining phase transition in the boundary theory \cite{Witten:1998zw}.

\begin{figure}[]
	\centering
	\includegraphics[scale = 0.35]{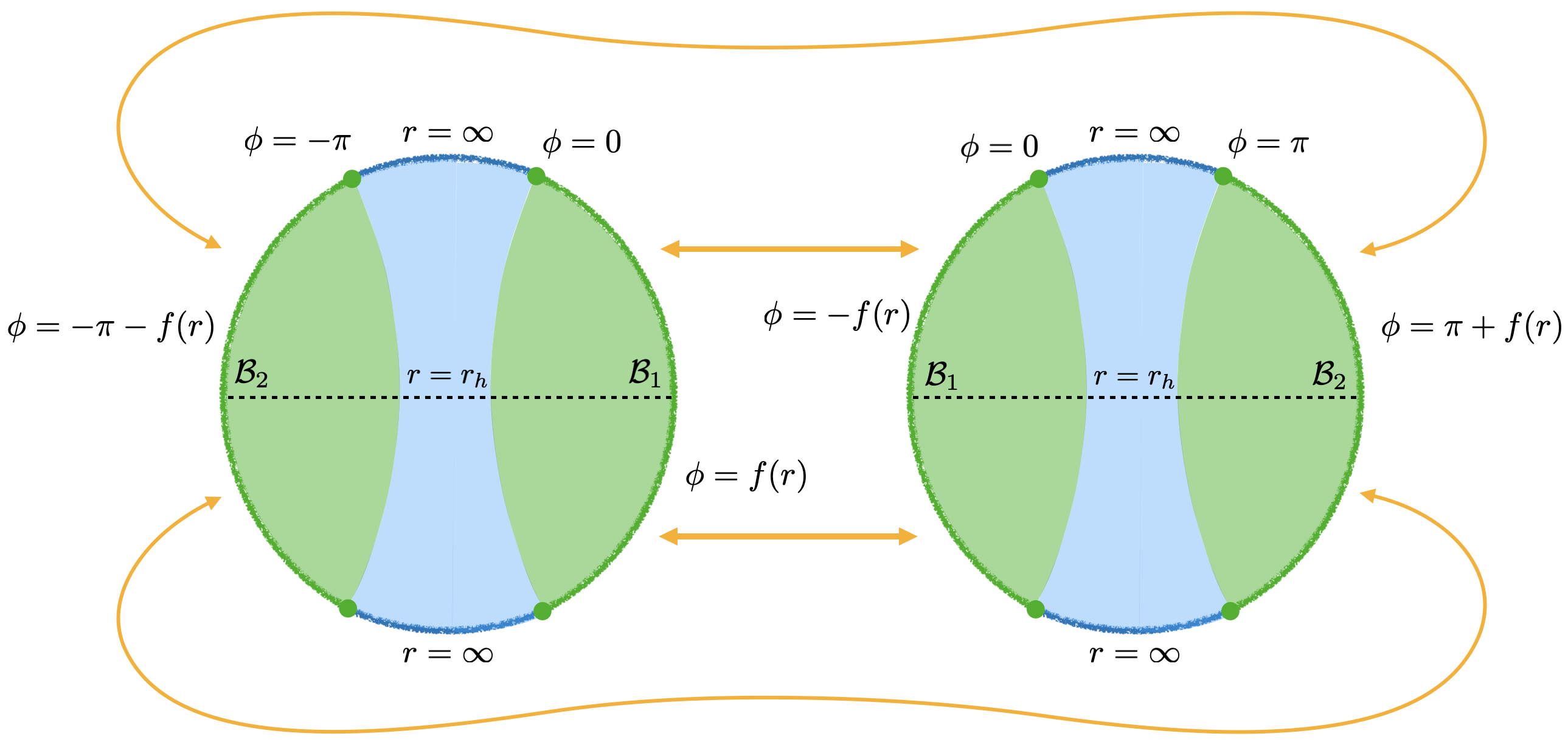}
	\caption{Sketch of a $t = 0$ slice of the geometry in the deconfined phase. The two branes, denoted with $\mathcal{B}_1$ and $\mathcal{B}_2$ in green, cross the wormhole at antipodal points on the boundary circle. The orange arrows indicate two geometries are glued together along the brane surfaces.}
	\label{fig:twobranes}
\end{figure}

We would like to examine the HP transition in the context of our defect construction as well. However, we must make a small modification here. Note that we expect that in the confining phase with a single defect, the background geometry would resemble thermal AdS$_3$ with a single two-dimensional brane wrapping the thermal circle and extending in the radial direction. The problem is that there would not be a smooth way to end the brane in the interior of the geometry and hence it must extend to reach the asymptotic boundary at two places. Hence this would describe a confining phase for the boundary CFT coupled to two defects. We therefore place two defects in the boundary CFT at $\phi = 0$ and $\phi = \pi$. The confining vacuum is then described by pure AdS$_3$ with a brane traversing the center of the space, \ie the solution studied in \cite{parti}. In the deconfined phase, the dual geometry is the BTZ black hole with two branes crossing the Einstein-Rosen bridge at antipodal points on the $\phi$ circle.\footnote{We note that similar constructions were considered in \cite{Germani:2006ea,BCFT} but using two $\mathbb{Z}_2$-orbifold branes. Further in \cite{Germani:2006ea}, the two branes collided at some fixed radius surface inside the horizon.} This geometry is sketched in figure \ref{fig:twobranes}.

In the following, we examine the black hole thermodynamics and the HP transition for this construction with two branes. We perform these calculations first for regular branes with usual worldvolume action in eq.~\reef{action}. In this case, the calculation of the HP transition was first sketched out in \cite{BCFT}. In this case, the dual boundary system was a two-dimensional CFT on a finite interval coupled to conformal boundaries at either end. However, the final expression for the critical temperature is identical to our result in eq.~\reef{gamble}. We then also repeat the calculations for branes supporting JT gravity, where the brane action is replaced by eq.~\reef{JTbrane1}. 
\subsection{Regular branes}

The total action of our system can be written schematically as follows
\begin{equation}
I = I_{\text{EH}} + I_{\text{GH}}^{\partial\mathcal{M}} + I_{\text{ct}} + I_{\mathcal{B}}  +2I_{\text{GH}}^{\mathcal{B}}\,.
\label{zoo}
\end{equation}
Here,  $I_{\text{EH}}$ is the usual Einstein-Hilbert action for the bulk, accompanied by the usual Gibbons-Hawking boundary term  $I_{\text{GH}}^{\partial\mathcal{M}}$ and the appropriate counterterm action $I_{\text{ct}}$ \cite{counterterms} on the regulator surfaces $\partial \mathcal{M}$ at $r=r_{\rm max}$, near the asymptotic boundaries on either side of the black hole. As usual, the counterterm action removes singularities coming from this surface and we send $r_{\rm max}\to \infty$ at the end of the calculation. Next, $I_{\mathcal{B}}$ is the standard worldvolume action for the branes $\B=\mathcal{B}_1\cup\mathcal{B}_2$, as shown in figure \ref{fig:twobranes}. 

We have also added Gibbons-Hawking terms on either side of the brane surfaces, \ie the final contribution $2I_{\text{GH}}^{\mathcal{B}}$.  This addition can be understood in two ways (\eg see discussion in \cite{partiii}): First, these boundary terms are required to define a proper variational problem if we treat the brane surface $\B$ as a boundary with Dirichlet boundary conditions $\delta g_{ij}=0$. From this perspective, the position of the brane is then fixed by the Israel junction conditions \reef{eq:junction}, rather than through the Einstein equations. Alternatively, if the branes were regulated with a finite thickness, the branes would produce large curvatures through the bulk Einstein equations, which in turn would contribute to the Einstein-Hilbert action. The latter would become a $\delta$-function contribution in the limit of zero thickness, but Israel's analysis \cite{Israel} shows that this contribution may be represented by excising the brane surface and evaluating the jump in the extrinsic curvature across this surface.

More explicitly, in Euclidean signature, the action \reef{zoo} reads
\begin{align}\label{eq:HPaction}
\begin{split}
I &= - \frac{1}{16\pi G_N} \int_{\mathcal{M}}\!\! \dd^3{x} \sqrt{g} \left(R + \frac{2}{L^2}\right)- \frac{1}{8\pi G_N}\int_{\partial \mathcal{M}}\!\!\!\! \dd^2x \sqrt{\gamma}\, K + \frac{1}{8\pi G_NL}\int_{\partial \mathcal{M}}\!\!\!\! \dd^2x  \sqrt{\gamma} \\ 
&\qquad+ T_0 \int_{\B_1\cup\B_2} \!\!\!\!\dd^2\sigma \sqrt{h}
- \frac{1}{8\pi G_N}\int_{\B_1\cup\B_2} \!\!\!\! \dd^2\sigma \sqrt{h}\, \Delta K\,,
\end{split}
\end{align}
where $\gamma_{ab}$ is the induced metric on asymptotic regulator surfaces and $K$ is the trace of the corresponding extrinsic curvature. Further, $h_{ab}$ is the induced metric on the brane surfaces and  $\Delta K$ is the jump in the trace of the extrinsic curvature across $\B=\B_1\cup\B_2$.  Notice that the latter provides a short-hand for the contributions of the Gibbons-Hawking terms on either side of the branes. Evaluating the action \eqref{eq:HPaction} onshell, we can simplify the expression to
\begin{align}
\begin{split}
I &= \frac{1}{4\pi G_NL^2} \int_{\mathcal{M}}\!\! \dd^3{x} \sqrt{g} - \frac{1}{8\pi G_N}\int_{\partial \mathcal{M}}\!\!\!\! \dd^2x \sqrt{\gamma}\left[ K - \frac{1}{L}\right]
\\ 
&\qquad+ \frac{1}{8\pi G_N} \int_{\B_1\cup\B_2} \!\!\!\!\dd^2\sigma \sqrt{h}\left[\frac{2}{L}\,\frac{k}{\sqrt{1+k^2}}
- \Delta K\right]\,,
\end{split}
\end{align}
where we have simplified the Einstein-Hilbert contribution using $R=-6/L^2$ from the Einstein equations, and we replaced the brane tension $T_0$ by an expression involving $k$ using eq.~\reef{eq:k}.  

\subsection*{BTZ computation} 
Our starting point is the Euclidean BTZ metric,
\begin{equation}
\dd s^2_{E} = \left(\frac{r^2}{L^2} - \mu^2\right)\frac{L^2}{R^2}\dd\tau^2 + \frac{\dd r^2}{\frac{r^2}{L^2}-\mu^2} + r^2 \dd\phi^2\,,
\end{equation}
where smoothness at $r=\mu L$ requires that $\tau$ is periodic with $\Delta\tau=1/T$ where $T = \frac{\mu}{2\pi R}$. On the boundary, $\phi$ has period $2\pi$ but as shown in figure \ref{fig:twobranes}, this periodicity is extended in the bulk by the presence of the two branes. In particular, following surfaces of constant $r$, we have $-f(r)<\phi<\pi+f(r)$ on the right side and  $-\pi-f(r)<\phi<f(r)$ on the left side.

The trace of the Einstein equations yields $R=-6/L^2$ and hence the on-shell Einstein-Hilbert term reduces to a simple integral over the spacetime volume
\begin{align}
\begin{split}
I_{\text{EH}} &= \frac{1}{4\pi G_N L^2} \int_{\mathcal{M}} \dd^3{x} \sqrt{g}\\
&=  \frac{1}{2\pi G_N L R} \int_0^{1/T} \!\!\!\dd\tau \int_{L\mu}^{r_{\text{max}}} \!\!\!r \dd r \int_{-\pi}^{0} \!\!\!\dd\phi + \frac{1}{\pi G_N L R} \int_0^{1/T} \!\!\!\dd\tau \int_{L\mu}^{r_{\text{max}}} \!\!\!r \dd r \int_{0}^{f(r)} \!\!\!\dd\phi \,, \end{split}
\end{align}
where the first term corresponds to the volume of the `pure' BTZ spacetime, while the second term is the added extra geometry by the presence of the two branes, where each enlarges the geometry by $f(r) = 1/\mu \arcsinh{(k\mu L/r)}$. In the above $r_{\text{max}}$ is some large radial cut-off. The integral then evaluates to 
\beqa\label{eq:I-EH}
I_{\text{EH}} &=& \frac{r_{\text{max}}^2}{4G_N T LR} + \frac{k\, r_{\text{max}}}{\pi G_N T R} \\
&&\qquad- \frac{L\,\mu}{2\pi G_NT R}\left(\sqrt{k^2(1+k^2)}+\frac{\pi \mu}{2}+ \arcsinh{k}\right) + \mathcal{O}\left(\frac{1}{r_{\text{max}}}\right)\,,\nonumber
\eeqa
which diverges for $r_{\text{max}} \to \infty$, which is to be expected. Notice that in the above integral we have accounted for both sides of the spacetime by using a symmetry argument and multiplying by an overall factor of 2. Let us now compute the following boundary term
\begin{equation}
I_{\partial\mathcal{M}} = - \frac{1}{8\pi G_N}\int_{\partial \mathcal{M}} \dd^2y \sqrt{\gamma} K_{\partial \mathcal{M}} \,,
\end{equation}
where $\gamma$ is the induced metric on $\partial\mathcal{M}: r = r_{\text{max}}$,
\begin{equation}
\dd s^2_{\gamma} = (r_{\text{max}}^2/L^2 - \mu^2)\frac{L^2}{R^2}\dd\tau^2 + r_{\text{max}}^2 d\phi^2\,,
\end{equation}
for some fixed $r = r_{\text{max}}$. The trace of the extrinsic curvature and the volume element on $\partial \mathcal{M}$ are
\begin{equation}
\sqrt{\gamma} = \frac{L r_{\text{max}}}{R}\sqrt{r_{\text{max}}^2/L^2 - \mu^2}, \;\;\; K = \frac{2r_{\text{max}}^2/L^2 - \mu^2}{r_{\text{max}}\sqrt{r_{\text{max}}^2/L^2 - \mu^2}}\,,
\end{equation}  so that
\begin{align}\label{eq:I-partial-M}
\begin{split}
I_{\partial\mathcal{M}} =\frac{\mu^2 - 2r_{\text{max}}^2/L^2}{4\pi G_N}\frac{L}{R}\left[\int_0^{1/T} \dd\tau \int_{-\pi}^{0}\dd{\phi} + 2\int_0^{1/T} \dd\tau \int_{0}^{f(r_{\text{max}})} \dd\phi \right]\\
=  \frac{\mu^2 L}{4G_N TR}- \frac{k}{\pi G_N TR} r_{\text{max}}  - \frac{1}{2G_NLR}r_{\text{max}}^2 + \mathcal{O}\left(\frac{1}{r_{\text{max}}}\right)\,.
\end{split}
\end{align}
The above linear divergence gets cancelled by the previous divergence in the Einstein-Hilbert action. Finally, for the branes
\begin{equation}
I_{\mathcal{B}} + 2I_{\text{GH}}^{\mathcal{B}}= \frac{1}{8\pi G_N} \int_{\B_1\cup\B_2} \!\!\!\!\dd^2\sigma \sqrt{h}\left[\frac{2}{L}\,\frac{k}{\sqrt{1+k^2}}
- \Delta K\right]\,,
\end{equation}
where $h_{ij}$ is the induced metric
\begin{equation}
\dd s^2_{\mathcal{B}} = \left(\frac{r^2}{L^2} -\mu^2\right)\frac{L^2}{R^2}\dd\tau^2 + \left(\frac{1}{\frac{r^2}{L^2} - \mu^2} + \frac{k^2}{\frac{r^2}{L^2} + k^2\mu^2} \right) \dd r^2.
\end{equation}
From the Israel's junction condition we easily get that $\Delta K = 2K_{\mathcal{B}} = \frac{4}{L} \sqrt{\frac{k^2}{k^2+1}}$. Furthermore, we have $\sqrt{h} = \frac{Lr}{R}\sqrt{\frac{1 + k^2}{r^2 + k^2L^2 \mu^2}}$, thus obtaining
\begin{align}\label{eq:I-GHB}
\begin{split}
I_{\mathcal{B}} + 2I_{\text{GH}}^{\mathcal{B}} &= - \frac{1}{2\pi G_NR}\int_{0}^{1/T} \dd \tau \int_{\mu L}^{r_{\text{max}}} r\sqrt{\frac{k^2}{r^2 + k^2L^2 \mu^2}} \dd r\\ 
&= \frac{\sqrt{k^2(1+k^2)}\mu L }{2\pi G_N RT}- \frac{k}{2\pi G_N RT}r_{\text{max}} + \mathcal{O}\left(\frac{1}{r_{\text{max}}}\right)\,.
\end{split}
\end{align}
Finally, we add the following counterterm
\begin{equation}\label{eq:I-ct}
I_{\text{ct}} =  \frac{1}{8\pi G_NL}\int_{\partial\mathcal{M}} \sqrt{\gamma}  = \frac{1}{4G_NLRT}r_{\text{max}}^2 + \frac{k}{2\pi G_NRT} r_{\text{max}} - \frac{\mu^2 L }{8G_NRT}+ \mathcal{O}\left(\frac{1}{r_{\text{max}}}\right)\,.
\end{equation}

Combining all of the contributions, we arrive at
\beqa
\lim_{r_{\text{max}} \to\infty} I_{E}& =& - \frac{1}{8G_N}\frac{L}{RT}\left[\mu^2 + \frac{4\mu}{\pi}\arcsinh{k}\right]
\nonumber\\
& =& - \frac{\pi^2}{3}\,c\,RT  - 2\,c_{\rm def}\,,\label{total}
\eeqa
where we have used $\mu=2\pi RT$, $c=3L/2G_N$ 
and $c_{\rm def}=c/3\,\arcsinh{k}$ to express the final result in terms of parameters of the boundary CFT. For branes of zero tension, \ie $k = 0$ and hence $c_{\rm def}=0$, the action reduces to that known for the BTZ black hole, \eg \cite{Eune_2013,Myung:2006sq, Maloney:2007ud} 
\begin{equation}
I_{\text{BTZ}} =  - \frac{\pi^2}{3}\,cRT \,.
\end{equation}
In the saddle-point approximation for the gravity path integral, this action \reef{total} yields the free energy as
\begin{equation}
F = T\,I_E =  - \frac{\pi^2}{3}\,c\,RT^2 - 2\,c_{\rm def}\,T\,.
\end{equation}
We can then apply standard thermodynamic identities to evaluate the energy and entropy of the system. The energy is given by
\beq
E=F-T\,\frac{\partial F}{\partial T}=\frac{\pi^2}{3}\,c RT^2\,,
\label{energyA}
\eeq
which we note is identical to the energy of the BTZ black hole without any branes. Similarly, the entropy is given by
\begin{align}
\begin{split}
S &= -\frac{\partial F}{\partial T}= \frac{2\pi^2}{3}\,c\,RT +  2\,c_{\rm def}\\
&= \frac{\pi L}{2G_N}\,R\mu + \frac{L}{G_N}\arcsinh{k}\\
&=\frac{1}{2G_N} \left[\int_{-\pi}^{0} r_{h}\,d\phi + 2\int_{0}^{f(r_h)} \!\!\!r_{h}\,d\phi\right]=\frac{A_{\rm horizon}}{4G_N}\,.
\end{split}
\label{sapp}
\end{align}
For the result in the first line, the first term is the expected thermal entropy matching that of the BTZ black hole, and the constant term is a temperature independent correction coming from the two defects.
In the final line with horizon radius $r_h=\mu L$ and from eq.~\reef{eq:Brane}, $f(r_h) = \frac1\mu\,\arcsinh{k}$, we
confirm that the entropy can also be evaluated with the usual Bekenstein-Hawking formula.

\subsection*{Thermal AdS computation}
The euclidean $AdS_3$ metric in global coordinate reads,
\begin{equation}\label{eq:metric-AdS}
\dd s^2_{E} = \left(\frac{r^2}{L^2} + 1\right)\frac{L^2}{R^2}\dd \tau^2 + \frac{\dd{r}^2}{\frac{r^2}{L^2} + 1} + r^2 \dd\phi^2\,,
\end{equation} 
with a brane profile
\begin{equation}
\phi = f(r) = \arcsin{\left(\frac{\sqrt{k}L}{r}\right)}.
\end{equation}
This can be derived from the brane in BTZ by setting $\mu^2 = -1$ and using the identity $-i\arcsinh(ix) = -\arcsin(-x)$. The Einstein-Hilbert term evaluates to
\begin{align}\label{eq:I-EH-thermal}
\begin{split}
I_{\text{EH}} &=  \frac{1}{2\pi G_NLRT}\int_{-(\pi + \epsilon)}^{\epsilon} \dd \phi \int_{0}^{r_{\text{max}}}r\dd{r} +  \frac{1}{2\pi G_NLRT}   \int_{\epsilon}^{\pi - \epsilon} \dd\phi \int_{0}^{\frac{kL}{\sin{\phi}}} r \dd r\\
&=\frac{1}{4G_NLRT}r_{\text{max}}^2 +\frac{k}{\pi G_NRT}r_{\text{max}} + \mathcal{O}\left(\frac{1}{r_{\rm max}}\right)\,,
\end{split}
\end{align}
where we have split the integration into two parts: one for the empty half-AdS space, and another for the extra geometry on the brane side. In the above, $\epsilon$ is a small angle that will be sent to zero. Because of this, we have used the relation $r_{\text{max}} \sim \frac{kL}{\epsilon}$ to obtain the asymptotic behaviour. The boundary term at infinity reads
\beq\label{eq:I-partial-M-thermal}
I_{\partial\mathcal{M}} =-\frac{2r_{\text{max}}^2+L^2}{4\pi G_NLRT} \left[ \int_{-\pi}^{0} \dd\phi + 2 \int_{0}^{\epsilon} \dd{\phi}\right] 
=- \frac{r_{\text{max}}^2}{2G_NLRT} - \frac{k\,r_{\rm max}}{\pi G_NRT} -\frac{ L}{4G_NRT} \,,
\eeq
where again we split into an empty AdS term, and a contribution coming from the surfaces $\partial \mathcal{M}$ entering the brane region and intersecting the brane for $\phi = \epsilon$. The brane term is
\begin{equation}
I_{\mathcal{B}} + 2I_{\text{GH}}^{\mathcal{B}} =  \frac{1}{8\pi G_N} \int_{\B} \dd^2\sigma \sqrt{h}\left[\frac{2}{L}\,\frac{k}{\sqrt{1+k^2}}
- \Delta K\right]\,,
\end{equation}
where now the integration goes over the single brane from $\phi = 0$ to $\phi = \pi$. More explicitly, the above integral reduces to
\beq\label{eq:I-GH-B-thermal}
I_{\mathcal{B}} + 2I_{\text{GH}}^{\mathcal{B}} = -  \frac{1}{2\pi G_N R T} \int_{kL}^{r_{\rm max}} \dd{r}\, r\sqrt{\frac{k^2}{r^2 - k^2L^2}} = -\frac{k\, r_{\rm max} }{2\pi G_{N} R T}+ \mathcal{O}\left( \frac{1}{r_{\rm max}}\right).
\eeq
Finally, we add the appropriate counterterm, obtaining
\begin{align}\label{eq:I-ct-thermal}
\begin{split}
I_{\text{ct}} &= \frac{r_{\text{max}}\sqrt{r_{\text{max}}^2+L^2 }}{4\pi G_NLRT}\left[ \int_{-\pi}^{0} \dd\phi + 2 \int_{0}^{\epsilon} \dd{\phi}\right]\\
&=  \frac{L}{8G_NRT} +  \frac{k}{2\pi G_NRT} r_{\rm max} + \frac{1}{4 G_NLRT}r_{\text{max}}^2 + \mathcal{O}\left(\frac{1}{r_{\rm max}}\right).
\end{split}
\end{align}
Putting everything together, we obtain
\begin{equation}
\lim_{r_{\rm max}\to \infty}\, I_{E} = -\frac{L}{8G_NRT}
=- \frac{c}{12RT}\,,
\end{equation}
which is the same result as in the empty AdS case. Hence we can write the free energy as
\beq
F = T\,I_E = - \frac{c}{12R}\,,
\eeq
which using the standard thermodynamic identities applied in eqs.~\reef{energyA} and \reef{sapp}, yields
\beq
E= - \frac{c}{12R} \,,\qquad\qquad S=0\,.
\eeq
As expected then, we have a Casimir energy associated with compactifying the boundary CFT on a circle but the entropy vanishes. We thus learn that the defect has no effect in the thermal AdS geometry. 

\subsection*{Critical temperature}
\begin{figure}[]
	\centering
	\includegraphics[scale = 0.4]{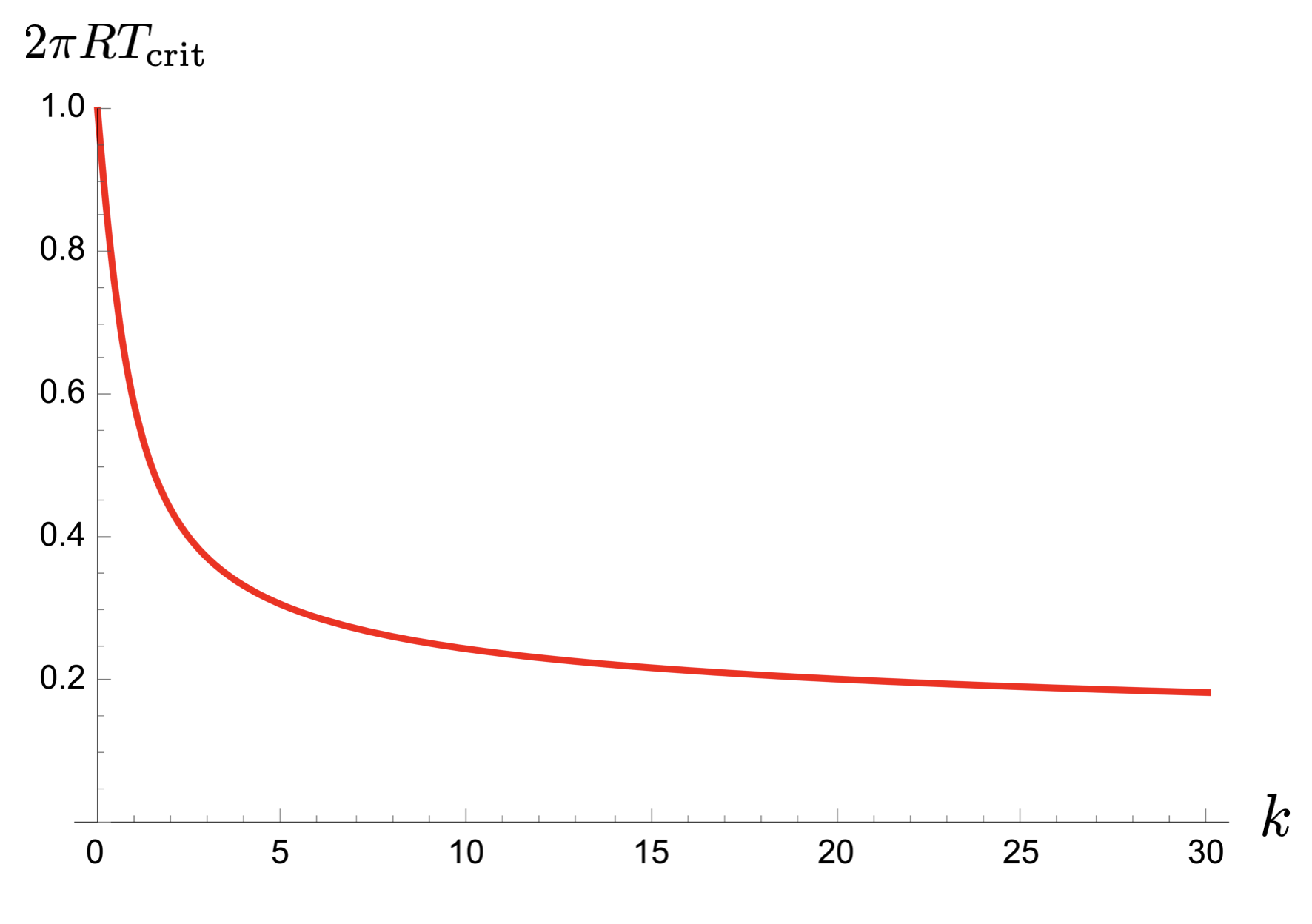}
	\caption{Plot of $2\pi R T_{\rm crit}$ versus $k$. The curve intersects the vertical axis at unity, as expected for the tensionless limit. For large tensions, the critical temperature tends to zero, as shown in eq.~\reef{gamble2}. }
	\label{fig:criticaltemp}
\end{figure}

Now that we have computed the on-shell actions for both geometries, the difference in free energies on shell is
\begin{align}
\Delta F = F_{BTZ} - F_{AdS_3} &= - \frac{L}{8G_NR}\left(\mu^2 + \frac{4\mu}{\pi}\arcsinh{k}\right)+ \frac{L}{8G_NR}\nonumber\\
&=- \frac{\pi^2}{3}\,c\,RT^2 - 2\,c_{\rm def}\,T +\frac{c}{12R}\label{label1}\,,
\end{align}
which vanishes when
\beq
T_{\rm crit} =  \frac{3}{\pi^2 R}\left( \sqrt{\frac{\pi^2}{36} + \frac{c_{\rm def}^2}{c^2}}-\frac{c_{\rm def}}{c}\right)
=  \frac{1}{\pi^2 R}\left( \sqrt{\frac{\pi^2}{4} + \arcsinh^2{k}}-\arcsinh{k}\right)\,,
\label{gamble}
\eeq
where the second expression comes using $c_{\rm def}/c =1/3\,\arcsinh{k}$. Hence the phase transition happens at this critical temperature. A plot of $T_{\rm crit}$ as a function of the variable $k$ is given in figure \ref{fig:criticaltemp}. For tensionless branes, \ie $k =0$, and the critical temperature reproduces  the well-known result $T_{\rm crit} = \frac{1}{2\pi R}$ for the BTZ black hole, \eg \cite{Eune_2013}.  As the tension increases towards the critical limit (and $k$ increases without bound as shown in eq.~\reef{eq:k}), the critical temperature decreases towards zero. Indeed, an expansion of the critical temperature \reef{gamble} for large $c_{\rm def}/c$ or large $k$ shows that this temperature vanishes as
\begin{equation}
T_{\rm crit}= \frac1{24R\,(c_{\rm def}/c)}+\cdots= \frac1{8R\,\log 2k} +\cdots\to 0\,.
\label{gamble2}
\end{equation}
As noted above, these results are in agreement with previous work in \cite{BCFT}, which studied the HP transition for a two-dimensional CFT on a finite interval coupled to conformal boundaries at either end. 

\subsection{JT branes}

We now study the HP transition for branes supporting dynamical JT gravity, as in appendix \ref{JTgrav}.  The computations follow those above, except that the brane action is replaced by that in eqs.~(\ref{JTbrane1}--\ref{JTbrane3}). We rewrite that brane action here for Euclidean signature 
\beqa
I'_{\rm brane} &=& I_{\rm JT} + I'_{\rm ct}
\label{JTbrane5}\\
&=& - \frac{1}{16\pi G_{\rm brane}} \int_{\B_1 \cup \B_2} \!\!\!\!\!\!\!\!\!\dd^2x \sqrt{h} \left[ \varphi_0 \tilde{R} + \varphi \left(\tilde{R} + \frac{2}{\ell_{\rm JT}^2} \right)\right]
+\frac{1}{4\pi G_{N}L}\int_{\B_1 \cup \B_2} 
 \!\!\!\!\!\!\!\!\dd^2 x \sqrt{h}\,.
\nonumber
\eeqa
As before, here $h$ is the determinant of the induced metric on the brane. 
Hence the total Euclidean action in eq.~\reef{zoo} is replaced by 
\beq\label{eq:JTEuclidean}
I = I_{\text{EH}} +  I_{\text{GH}}^{\partial\mathcal{M}} + I_{\text{ct}}+I'_{\B}+2I_{\text{GH}}^{\mathcal{B}} + I_{\text{GH}}^{\partial\mathcal{B}} + I_{\rm ct}^{\partial{\cal B}} \,.
\eeq
Note that since the brane action $I'_{\B}$ given in eq.~\reef{JTbrane5} involves a dynamical gravity theory, we have added a Gibbons-Hawking boundary term $I_{\text{GH}}^{\partial\mathcal{B}}$ on the branes' boundary $\partial \mathcal{B}$, \ie the intersection of the two branes $\mathcal{B}_1$ and $\mathcal{B}_2$ with the asymptotic boundary $\partial \mathcal{M}$. Further, we must also add an additional counterterm $I_{\rm ct}^{\partial{\cal B}}$ on the boundary of the branes, \eg \cite{Nayak:2018qej,Harlow:2018tqv,Witten:2020ert}. We note that  we only need to compute $I'_{\B}+ I_{\text{GH}}^{\partial\mathcal{B}}+I_{\rm ct}^{\partial{\cal B}}$ in the following, since the rest of the terms have been computed in the previous section and their results remain unchanged here. 

\subsection*{BTZ computation} 
We begin by evaluating the action for the black hole background. In Euclidean signature, the induced metric \reef{eq:BH1} on the brane becomes
\begin{equation}
\dd s^2_{\mathcal{B}} = \left(\frac{r^2}{L^2} -\mu^2\right)\frac{L^2}{R^2}\,\dd\tau^2 + \left(\frac{1}{\frac{r^2}{L^2} - \mu^2} + \frac{k^2}{\frac{r^2}{L^2} + k^2\mu^2} \right) \dd r^2.
\end{equation}
and so we have $\sqrt{h} = \frac{Lr}{R}\sqrt{\frac{1 + k^2}{r^2 + k^2L^2 \mu^2}}$. As noted above with dynamical gravity on the branes, we need to equip the brane action with an appropriate Gibbons-Hawking boundary term 
\beq
I_{\rm GH}^{\partial \mathcal{B}} = - \frac{1}{8\pi G_{\rm brane}} \int_{\partial \mathcal{B}} \dd \tau \sqrt{\sigma} \left(\varphi_0 + \varphi_{b}  \right)K_{\partial \mathcal{B}}\,,
\label{JTbound1}
\eeq
where $\varphi_{b}$ is the value of the dilaton at the boundary, and $\sigma$ and $K_{\partial \mathcal{B}}$ are the determinant of the one-dimensional induced metric and the extrinsic curvature of the two branes' boundary $\partial \mathcal{B}$, to wit,
\beq
\dd s^2_{\partial \mathcal{B}} = \left(\frac{r_{\rm max}^2}{L^2} -\mu^2\right)\frac{L^2}{R^2} \dd\tau^2,\quad \sqrt{\sigma} = \sqrt{\frac{r_{\rm max}^2}{L^2} -\mu^2} \frac{L}{R} , \quad K_{\partial\mathcal{B}} = \frac{1}{L}\sqrt{\frac{r_{\rm max}^2 + k^2 \mu^2 L^2}{(1 + k^2)(r_{\rm max}^2 - \mu^2 L^2)}}\,.
\label{cornball1}
\eeq
Further, we include a counter-term $I_{\rm ct}^{\partial{\cal B}}$ which ensures the on-shell action is finite
\beq
I_{\rm ct}^{\partial{\cal B}} = \frac{1}{8\pi G_{\rm brane}\ell_{\rm JT}} \int_{\partial \mathcal{B} } \dd \tau \sqrt{\sigma} \,\varphi_{b}\,.
\label{cornball2}
\eeq

On shell, we have 
\beq\label{label0}
\tilde{R} = - 2/\ell_{\rm JT}^2 \quad \text{and} \quad \varphi = \frac{\bar{\varphi}_r}{\ell_{\mathcal{B}}\mu}\rho +\frac{G_{\rm brane}}{G_{\rm eff}}\frac{\ell_{\rm JT}^2}{\hat{\ell}_{\rm eff}^2}.
\eeq
where $\rho = \sqrt{r^2 + k^2\mu^2 L^2}$. Hence, the term $I_{\B}$ becomes a simple integral over the branes' volume
\beq
I'_{\B}  =  \frac{\varphi_0}{4\pi G_{\rm brane} \ell_{\rm JT}^2} \int_{\B_1 \cup \B_2}\!\!\!\!\!\! \dd^2x\, \sqrt{h} + \frac{1}{2\pi G_{N}L}\int_{\B_1 \cup \B_2}\!\!\!\!\!\! \dd^2x\, \sqrt{h}\,.
\eeq
which evaluates to
\begin{align}\label{eq:IB}
\begin{split}
I'_{\B}  &=   \left(\sqrt{(1+k^2)(r_{\rm max}^2 + k^2 L^2 \mu^2)} - (1+k^2)\mu L\right)\left(\frac{\varphi_0 L }{4\pi G_{\rm brane} \ell_{\rm JT}^2 R T} + \frac{1}{2\pi G_{N} R T}\right)\\
&=  \frac{\varphi_0}{4\pi G_{\rm brane} \ell_{\rm JT}R T} \, r_{\rm max} + \frac{\ell_{\rm JT}}{2\pi G_{N} LRT}\, r_{\rm max} - \frac{\mu \varphi_0}{4\pi G_{\rm brane} R T} - \frac{\mu \ell_{\rm JT}^2}{2\pi G_{N} L RT} + \mathcal{O}\left(\frac{1}{r_{\rm max}}\right).
\end{split}
\end{align}
In simplifying the above we have used the relation $1 + k^2 = \ell_{\rm JT}^2/L^2$. 

We are then left to compute contributions of the corresponding Gibbons-Hawking term \reef{cornball1} and counterterm action \reef{cornball2}. This yields
\begin{align}\label{eq:IGHbd}
\begin{split}
I_{\rm GH}^{\partial \mathcal{B}}+I_{\rm ct}^{\partial{\cal B}} 
&= - \frac{1}{4\pi G_{\rm brane} L R T }\sqrt{\frac{r^2_{\rm max} + k^2 \mu^2 L^2 }{1 + k^2}} (\varphi_0+\varphi_b) + \frac{L \varphi_b}{4\pi G_{\rm brane}\ell_{\rm JT} RT }\sqrt{\frac{r_{\rm max}^2}{L^2} - \mu^2}\\
&= -\frac{\varphi_0}{4\pi G_{\rm brane} \ell_{\rm JT } R T }\, r_{\rm max} - \frac{L}{4\pi G_N \ell_{\rm JT} RT}\, r_{\rm max}- \frac{\bar{\varphi}_r}{4\pi G_{\rm brane}\ell_{JT} R T}\frac{\rho_{\rm max}^2}{\mu \ell_{\B}}\\
&+ \frac{L}{4 \pi G_{N} \ell_{\rm JT}  RT}\, r_{\rm max} + \frac{\bar{\varphi}_r}{4\pi G_{\rm brane} \mu \ell_{\rm JT}^2 RT} \, r_{\rm max}^2 + \mathcal{O}\left(\frac{1}{r_{\rm max}}\right)\,.
\end{split}
\end{align}
In the above, we have used the relations $G_{\rm eff} = G_{N}/L$ and $L^2/\ell_{\rm eff}^2 \approx L^2/\ell_{\rm JT}^2$ (valid in a large $k$ expansion). 

Now combining eqs.~\reef{eq:IB} and \reef{eq:IGHbd} together, we obtain
\beqs
I'_{\B}  + I_{\rm GH}^{\partial \mathcal{B}} +I_{\rm ct}^{\partial{\cal B}} &=  \frac{\ell_{\rm JT}}{2\pi G_{N}L RT}\, r_{\rm max} - \frac{\mu \varphi_0}{4\pi G_{\rm brane} R T} - \frac{ \mu \ell_{\rm JT}^2}{2\pi G_{N}LRT} - \frac{\bar{\varphi}_r\,k^2 \mu L^2}{4\pi G_{\rm brane}\ell^2_{JT} R T}\,,
\eeqs
which, in the large $k$ limit, simplifies to
\beq
I'_{\B}  + I_{\rm GH}^{\partial \mathcal{B}} +I_{\rm ct}^{\partial{\cal B}} =   - \frac{(\varphi_0 + \bar{\varphi}_r)\mu}{4\pi G_{\rm brane} R T} - \frac{ \mu \ell_{\rm JT}^2}{2\pi G_{N}LRT} + \frac{k}{2\pi G_{N}RT}\, r_{\rm max}\,.
\eeq

We are now ready to add the remaining contributions for the total action: $I_{\rm EH} + 2 I^{\B}_{\rm GH}+ I^{\partial \mathcal{M}}_{\rm GH} + I_{\rm ct}$, which can be found in eqs.~\eqref{eq:I-EH}, \eqref{eq:I-partial-M}, \eqref{eq:I-GHB} and \eqref{eq:I-ct}. The final result becomes
\beq\label{eq:I-JT}
I_E =  - \frac{\mu^2 L}{8G_N RT} - \frac{\mu L}{2\pi G_N RT}\arcsinh{k} - \frac{( \varphi_0+\bar{\varphi}_r)\mu}{4\pi G_{\rm brane} R T}\,.
\eeq
The free energy is given by
\beqs
F = T I_E &= -\frac{\mu^2 L}{8 G_N R}-\frac{\mu L}{2\pi G_N R}\arcsinh{k} - \frac{(\varphi_0+\bar{\varphi}_r)\mu}{4\pi G_{\rm brane} R } \\
&= - \frac{\pi^2}{3}\,c\,RT^2 - 2\,c'_{\rm def}\,T\,,
\eeqs
where the defect central charge is increased by the presence of JT gravity with
\beq
c'_{\rm def} = \frac{c}3\,\arcsinh k + \frac{\varphi_0+\bar{\varphi}_r}{ 4G_{\rm brane}}\,.
\label{cdef88}
\eeq
We might identify $c_{\rm JT} = \frac{\varphi_0+\bar{\varphi}_r}{ 4G_{\rm brane}}$ as the central charge of the degrees of freedom dual to JT gravity. 
The energy is now given by
\beq
E = F- T \frac{\partial F}{\partial T} = \frac{\pi^2}{3}\,c \, R T^2\,,
\label{ener88}
\eeq
while the entropy becomes 
\beqs
S &= -\frac{\partial F}{\partial T}  = \frac{2\pi^2}{3}\,c\,RT +  2\,c'_{\rm def}\\
&= \frac{A_{\rm horizon}}{4G_N} + 2\, \frac{\varphi_0+\varphi}{4G_{\rm brane}} \Big|_{\rm horizon}\,,
\label{entro88}
\eeqs
where the factor of 2 in front of the dilaton contribution arises because there are two branes in the present calculation.

\subsection*{Thermal AdS computation} 

We now evaluate the on-shell action in the thermal AdS case, with metric as in eq.~\eqref{eq:metric-AdS}. Once again, we only need to compute $I'_{\cal B} + I_{\rm GH}^{\partial{\cal B}} + I_{\rm ct}^{\partial {\cal B}}$, since the rest of the terms are similar to the non JT gravity case.

For global AdS$_2$, the dilaton profile is given by the constant term only\footnote{The linear term proportional to $\bvphi_r$ is ruled out here because of the boundary condition for global AdS$_2$. That is, the asymptotic dilaton must be constant (\ie independent of $\tau$) along the surface $r=r_{\rm max}$ in the coordinates given in  eq.~\eqref{eq:metric-AdS}.}
\beq
\varphi = \frac{G_{\rm brane}}{G_{\rm eff}} \frac{\ell_{\rm JT}^2}{\hat{\ell}_{\rm eff}^2}\,,
\eeq
and $\tilde{R} = -2/\ell_{\rm JT}^2$. 

We begin by evaluating 
\beqs
I'_{\cal B} & = \left(\frac{\varphi_0}{4 \pi G_{\rm brane} \ell_{\rm JT}^2} + \frac{1}{2\pi G_N L}\right) \int d^2x \sqrt{h}\\
& =  \left(\frac{\varphi_0}{4 \pi G_{\rm brane} \ell_{\rm JT}^2} + \frac{1}{2\pi G_N L}\right) \int_0^{\frac{1}{T}} d\tau \int_{kL}^{r_{\rm max}} dr \frac{Lr}{R}\sqrt{\frac{1+k^2}{r^2-k^2L^2}}\\
%
%
& =\left(\frac{\varphi_0}{4 \pi G_{\rm brane} \ell_{\rm JT}RT} + \frac{\ell_{\rm JT}}{2\pi G_N L RT}\right) \, r_{\rm max} + {\cal O}\left(\frac{1}{r_{\rm max}}\right)\,,
\eeqs
and 
\beqs
I_{\rm GH}^{\partial \mathcal{B}}+I_{\rm ct}^{\partial{\cal B}} 
&= - \frac{1}{4\pi G_{\rm brane} L R T }\sqrt{\frac{r^2_{\rm max} - k^2 L^2 }{1 + k^2}} (\varphi_0+\varphi_b) + \frac{L \varphi_b}{4\pi G_{\rm brane}\ell_{\rm JT} RT }\sqrt{\frac{r_{\rm max}^2}{L^2} +1}\\
&= -\frac{\varphi_0}{4\pi G_{\rm brane} \ell_{\rm JT } R T }\, r_{\rm max} + \mathcal{O}\left(\frac{1}{r_{\rm max}}\right)\,.
\eeqs
Combining these results together, we find
\beqs
I'_{\B}  + I_{\rm GH}^{\partial \mathcal{B}} +I_{\rm ct}^{\partial{\cal B}} &=  \frac{\ell_{\rm JT}}{2\pi G_{N}L RT}\, r_{\rm max}\,,
\eeqs
which, in the large $k$ limit, simplifies to
\beq
I'_{\B}  + I_{\rm GH}^{\partial \mathcal{B}} +I_{\rm ct}^{\partial{\cal B}} =    \frac{k}{2\pi G_{N}RT}\, r_{\rm max}.
\eeq
Lastly, adding the remaining contributions:  $I_{\rm EH} + 2 I^{\B}_{\rm GH}+ I^{\partial \mathcal{M}}_{\rm GH} + I_{\rm ct}$, which can be found in eqs.~\eqref{eq:I-EH-thermal}, \eqref{eq:I-partial-M-thermal}, \eqref{eq:I-GH-B-thermal} and \eqref{eq:I-ct-thermal}, we find the same answer as the case without JT gravity
\beq
\lim_{r_{\rm max}\to \infty} I_E = -\frac{L}{8G_N RT}   = -\frac{c}{12 RT}\,,
\eeq
and so the free energy, energy and entropy are the same as well
\beq
F = T I_E =  -\frac{c}{12 R} \,, \quad
E =  -\frac{c}{12 R} \,, \quad S = 0\,.
\eeq
\subsection*{Critical temperature} 

The difference of the free energies is the same as  without JT gravity, except the central charge of the conformal defect receives the dilaton contribution $c_{\rm def} \to c'_{\rm def} = c_{\rm def} + c_{\rm JT} =\frac{c}3\,\arcsinh{k}+ \frac{\varphi_0+\bvphi_r}{4G_{\rm brane}}$. Hence eq.~\reef{label1} is replaced with
\beq
\Delta F = - \frac{\pi^2}{3}\,c\,RT^2 - 2\,c'_{\rm def}\,T +  \frac{c}{12 R} \,.
\eeq
Then the new critical temperature is
\beqs
T_{\rm crit} &=  \frac{3}{\pi^2 R}\left( \sqrt{\frac{\pi^2}{36} + \left(\frac{c'_{\rm def}}{c}\right)^2}-\frac{c'_{\rm def}}{c}\right)\,,\\
&=  \frac{1}{\pi^2 R}\left( \sqrt{\frac{\pi^2}{4} + \left(\arcsinh{k}+\frac{3}{4c} \frac{\varphi_0+\bvphi_r}{G_{\rm brane}}\right)^2}-\arcsinh{k}-\frac{3}{4c} \frac{\varphi_0+\bvphi_r}{G_{\rm brane}}\right)\,,
\eeqs
and in the limit where the ratio $c'_{\rm def}/c$ is large, we find
\begin{equation}
T_{\rm crit}= \frac1{24R\,(c'_{\rm def}/c)}+\cdots\to 0\,.
\label{gamble88}
\end{equation}

\bibliographystyle{jhep}
\bibliography{references}


\end{document}